\documentclass[12pt]{article}
\def\appendix{
\setcounter{section}{0}
\renewcommand{\thesection}{Appendix \Alph{section}:}
\renewcommand{\thesubsection}{\Alph{section}.\arabic{subsection}.}
\renewcommand{\theequation}{\Alph{section}.\arabic{equation}}
}
\def\section#1{
    \addtocounter{section}{1}\setcounter{subsection}{0}\setcounter{equation}{0}
    \vskip8mm\begin{center}{\bf\thesection~#1}\end{center}}
\def\subsection#1{\addtocounter{subsection}{1}
    \vskip6mm\noindent{\sc\thesubsection~#1}\vskip4mm}
\def\paragraph#1{
    \vskip6mm\noindent{\it #1}\par\vskip3mm}

\newcommand{\vev}[1]{\left<{#1}\right>}
\newcommand{\bra}[1]{\left<{#1}\right|}
\newcommand{\ket}[1]{\left|{#1}\right>}
\newcommand{\cbra}[1]{_{\rm C}\!\!\left<\!\left<{#1}\right.\right|}
\newcommand{\cket}[1]{\left|\left.{#1}\right>\!\right>_{\rm \!C}}
\newcommand{\ibra}[1]{_{\rm I}\!\!\left<\!\left<{#1}\right.\right|}
\newcommand{\iket}[1]{\left|\left.{#1}\right>\!\right>_{\rm \!I}}
\newcommand{\tprod}{{\textstyle\prod}}

\newcommand{\mint}{{\displaystyle\int}}
\newcommand{\tfrac}[2]{{\textstyle\frac{#1}{#2}}}
\newcommand{\mfrac}[2]{{\displaystyle\frac{#1}{#2}}}
\newcommand{\Up}{{\Upsilon}}
\newcommand{\NS}{{\rm NS}}
\newcommand{\R}{{\rm R}}
\newcommand{\cF}{{\cal F}}
\newcommand{\cG}{{\cal G}}
\newcommand{\bG}{{\bf G}}
\newcommand{\bS}{{\bf S}}

\newcommand{\ch}{{\hat{c}}}
\newcommand{\ep}{{\epsilon}}
\newcommand{\bep}{{\bar{\epsilon}}}
\newcommand{\mep}{{\underline{\epsilon}}}
\newcommand{\btau}{{\bar{\tau}}}

\begin{document}

\renewcommand{\thefootnote}{\fnsymbol{footnote}}
\setcounter{footnote}{0}
\setcounter{section}{0}
\baselineskip = 0.6cm
\pagestyle{empty}


\baselineskip 5mm
\hfill\vbox{\hbox{YITP-02-07}
            \hbox{hep-th/0202032} }

\baselineskip0.8cm\vskip2cm

\begin{center}
 {\large\bf Super Liouville Theory with Boundary}
\end{center}

\vskip10mm

\baselineskip0.6cm
\begin{center}
       Takeshi Fukuda\footnote{ \tt tfukuda@yukawa.kyoto-u.ac.jp}
   and Kazuo Hosomichi\footnote{\tt hosomiti@yukawa.kyoto-u.ac.jp }
 \\ \vskip2mm
{\it Yukawa Institute for Theoretical Physics \\
     Kyoto University, Kyoto 606-8502, Japan} \vskip3mm
\end{center}

\vskip8mm\baselineskip=3.5ex
\begin{center}{\bf Abstract}\end{center}\par\smallskip

   We study $N=1$ super Liouville theory on worldsheets
 with and without boundary.
 Some basic correlation functions on a sphere or a disc
 are obtained using the properties of degenerate representations
 of superconformal algebra.
 Boundary states are classified by using the modular
 transformation property of annulus partition functions,
 but there are some of those whose wave functions
 cannot be obtained from the analysis of modular property.
 There are two ways of putting boundary condition on supercurrent,
 and it turns out that the two choices lead to different
 boundary states in quality.
 Some properties of boundary vertex operators are also presented.
 The boundary degenerate operators are shown to connect two
 boundary states in a way slightly complicated than the bosonic
 case.

\vspace*{\fill}
\noindent February~~2002
\setcounter{page}{0}
\newpage


\setcounter{footnote}{0}
\setcounter{section}{0}
\pagestyle{plain}
\renewcommand{\thesection}{\arabic{section}.}
\renewcommand{\thesubsection}{\arabic{section}.\arabic{subsection}.}
\renewcommand{\theequation}{\arabic{section}.\arabic{equation}}
\renewcommand{\thefootnote}{\arabic{footnote}}
\setcounter{footnote}{0}


\section{Introduction}

   Conformal field theories on worldsheets with boundary 
 play an important role in understanding some aspects of
 string theory, since it gives a worldsheet description
 of D-branes.
 Conformal field theories in general have large symmetry
 which includes Virasoro symmetry, and they are generated by
 holomorphic currents.
 On worldsheets without boundary there are two copies of the same
 symmetry algebra corresponding to the left- and the right-moving
 sectors, and on the boundary of worldsheets the two are related to
 each other by certain boundary condition which preserves one
 copy of the symmetry algebra.
 From the representation theoretical point of view the classification
 of {\it boundary states} reduces to that of possible boundary
 conditions and their solutions.

   On the other hand, some conformal field theories are endowed
 with a Lagrangian, and it is also available even for worldsheets with
 boundary if suitable boundary terms are incorporated.
 From this viewpoint, the classification of boundary states
 corresponds to that of possible boundary terms along with the
 boundary conditions on the fields.

   These two viewpoints have been shown to be consistent in
 \cite{FZZ,ZZ} for Liouville theory.
 The boundary states are classified in a complete way through
 the analysis of modular property of annulus partition functions,
 and some boundary states correspond to the addition of a
 boundary interaction term with certain values of coupling constant.
 It was also the first case where the classification of Cardy states
 was done for non-compact CFTs having continuous spectrum of
 representations.
 Based on this idea the Liouville theory with boundary has been
 analyzed in \cite{H,PT}, and the analysis of boundary states has
 also been made recently for more involved CFTs such as CFT
 on Euclidean $AdS_3$\cite{GKS,PS,LOP,PST}.

   In this paper we consider the $N=1$ supersymmetric extension
 of Liouville theory in the presence of boundary,
 using the techniques developed in \cite{FZZ,ZZ}.
 This theory has been analyzed for decades and some old references
 include \cite{Arvis,DHoker,Babelon,ZP,ACDH,dFK,AD,DA}.
 One will face some complexity due to the presence of
 NS and R sectors, and a careful analysis reveals
 what kind of new features arises as a result
 of supersymmetrization.
 For the case without boundary, the exact results for basic correlators
 has been obtained in \cite{Poghosian, RS}.
 There has also been some recent works on the case with boundary\cite{ARS}.
 Since super Liouville theory is one of the simplest CFTs
 with $N=1$ worldsheet supersymmetry, our result should contain
 many of the properties which all the $N=1$ supersymmetric
 CFTs have in common.

~

   This paper is organized as follows.
 In section 2 we analyze the $N=1$ super Liouville theory without
 boundary, especially on a sphere.
 The calculation of basic correlation functions which has been done
 in  \cite{Poghosian,RS} are reviewed.
 We first summarize the spectrum of degenerate representations
 in $N=1$ superconformal algebra.
 Then we calculate various structure constants using the most
 fundamental degenerate operators which will be denoted as
 $\Theta_{-b/2}^{\ep\bep}$, under the reasonable assumption
 that the product of them with any operators are expanded
 into two discrete terms.
 The consistency of this assumption is investigated by solving
 the differential equation for four-point functions containing
 $\Theta_{-b/2}^{\ep\bep}$.
 In section 3 we analyze the theory on worldsheets with boundary,
 especially on a disc.
 The modular property of annulus partition functions are investigated,
 from which we obtain the wave functions for some Cardy states.
 There are some others whose wave functions cannot be determined from
 the analysis of modular property, and we determine them
 through the analysis of disc one-point functions.
 The two-point functions of boundary operators are also obtained.
 The results for reflection coefficients are consistent with the 
 argument of density of open string states, but it turns out
 that the reflection coefficients differ for each operator in a
 single supermultiplet.
 The last section gives a brief summary of our results and some
 discussions.

\section{${\cal N}=1$ Super Liouville Theory}

   The supersymmetric extension of Liouville theory was found
 in \cite{Polyakov2}.
 It is described by a boson $\phi$ and its superpartner $\psi$,
 and the action on flat worldsheet reads
\begin{equation}
  I = \frac{1}{2\pi}\int d^2zd\bar{\theta}d\theta D\Phi\bar{D}\Phi
     +2i\mu\int d^2zd\bar{\theta}d\theta e^{b\Phi},
\end{equation}
 where we employed the superfield formalism
\begin{equation}
  \Phi= \phi+i\theta\psi+i\bar{\theta}\bar{\psi}+i\theta\bar{\theta}F,~~~
  D=\partial_\theta+\theta\partial_z,~~~
  \bar{D}=\partial_{\bar{\theta}}+\bar{\theta}\partial_{\bar{z}}.
\end{equation}
 The reader should note that there is a linear dilaton coupling
 hidden in the action.
 In \cite{DHK} it was analyzed as a two-dimensional theory of
 supergravity with superconformal symmetry.
 Superfield expression for linear dilaton coupling
 can be found in \cite{CKT}.

   We regard this theory as a free CFT of $\phi$ and $\psi$ with
 a linear dilaton coupling,
\begin{equation}
  I = \frac{1}{2\pi}\int d^2z\left[
       \partial\phi\bar{\partial}\phi+\tfrac{QR\phi}{4}
      +\psi\bar{\partial}\psi+\bar{\psi}\partial\bar{\psi}-F^2
      \right],~~~
      Q=b+b^{-1}
\end{equation}
 perturbed by the following interaction
\begin{equation}
   2i\mu\int d^2z
   \left[ibFe^{b\phi}+b^2\psi\bar{\psi}e^{b\phi}\right].
\end{equation}
 In what follows we shall neglect the auxiliary field $F$ which
 yields a contact interaction, assuming the analyticity of correlators
 or OPEs that allow us to calculate any quantity by the continuation
 from the region where contact interactions can be neglected.
 See \cite{GS, DS, dFK} for more detailed argument on this point.
 Thus we shall treat the super Liouville theory as the free CFT
 of $\phi$ and $\psi$ perturbed by
\begin{equation}
  S_{\rm int} \equiv 2i\mu b^2\int d^2z \psi\bar{\psi}e^{b\phi}.
\end{equation}

   The stress tensor $T$ and the supercurrent $T_F$ of the free
 theory are given by the Feigin-Fuchs representation
\begin{equation}
\begin{array}{rcl}
  T   &=& -\frac{1}{2}(\partial\phi\partial\phi
                  -Q\partial^2\phi+\psi\partial\psi),\\
  T_F &=& i(\psi\partial\phi - Q\partial\psi).
\end{array}
\end{equation}
 They satisfy the super Virasoro algebra with
 $c=\tfrac{3\ch}{2}=\tfrac{3}{2}(1+2Q^2)$
\begin{eqnarray}
  T(z)  T(0)  &\sim&
  \frac{3\ch}{4z^4}+\frac{2T(0)}{z^2}+\frac{\partial T(0)}{z}, \nonumber \\
  T(z)  T_F(0)&\sim&\frac{3T_F(0)}{2z^2}+\frac{\partial T_F(0)}{z},\\
  T_F(z)T_F(0)&\sim&\frac{\ch}{z^3}+\frac{2T(0)}{z}. \nonumber
\end{eqnarray}
 We shall concentrate on the left-moving sector for the time being,
 in order to clarify the symmetry structure of the theory. 
 We work with the primaries $V_\alpha=e^{\alpha\phi}$
 and their superpartners $\Lambda_\alpha=-i\alpha\psi e^{\alpha\phi}$,
 which satisfy
\begin{equation}
\begin{array}{rcl}
  T(z)V_\alpha(0)&\sim&
   \mfrac{h_\alpha V_\alpha(0)}{z^2}
  +\mfrac{\partial V_\alpha(0)}{z}, \\
  T(z)\Lambda_\alpha(0)&\sim&
   \mfrac{(h_\alpha+\frac{1}{2}) \Lambda_\alpha(0)}{z^2}
  +\mfrac{\partial \Lambda_\alpha(0)}{z}, \\
  T_F(z)V_\alpha(0)&\sim&
   \mfrac{\Lambda_\alpha(0)}{z}, \\
  T_F(z)\Lambda_\alpha(0)&\sim&
   \mfrac{2h_\alpha V_\alpha(0)}{z^2}
  +\mfrac{\partial V_\alpha(0)}{z},
\end{array}
\end{equation}
 with $h_\alpha=\alpha(Q-\alpha)/2$.
 They are NS vertices and correspond to space-time bosons.
 We also consider the R vertices corresponding to space-time
 fermions, which are given by spin fields
 $\Theta_\alpha^\pm=\sigma^\pm e^{\alpha\phi}$.
 They obey the following transformation property
\begin{equation}
\begin{array}{rcl}
  T(z)\Theta_\alpha^\pm(0) &\sim&
  \mfrac{(h_\alpha+\frac{1}{16})\Theta_\alpha^\pm(0)}{z^2}
 +\mfrac{\partial\Theta_\alpha^\pm(0)}{z} , \\
  T_F(z)\Theta_\alpha^\pm(0) &\sim&
  \mfrac{p_\alpha\Theta_\alpha^\mp(0)}{\sqrt{2}z^{3/2}}+\cdots,~~~~
  p_\alpha =\mfrac{i(Q-2\alpha)}{2}.
\end{array}
\end{equation}
 The most important property of spin fields is that the
 supercurrent $T_F$ becomes double-valued around them.
 The spin field $\sigma^\pm$ are defined to satisfy
\begin{equation}
  \psi(z)\sigma^\pm(0)\sim \frac{\sigma^\mp(0)}{\sqrt{2}z^{1/2}}.
\label{pxs}
\end{equation}

   We can analyze the theory perturbatively, by expanding
 any quantity as a power series in the cosmological constant.
 However, due to the momentum conservation in Linear dilaton theory, 
 any correlators of operators of definite Liouville momentum
 have only one contribution from a specific order of $\mu$.
 This can easily be seen by employing the path integration approach
 and perform the integration over the zero-mode of $\phi$
 as discussed in \cite{GL}.
 Then we find, for example,
\begin{equation}
  \vev{\tprod_iV_{\alpha_i}(z_i)}
 =b^{-1}\Gamma(-N)\vev{\tprod_iV_{\alpha_i}(z_i)S_{\rm int}^N}_{\rm Wick}
\label{GLI}
\end{equation}
 where the suffix ``Wick'' represents the ordinary Wick contraction
 with respect to free fields and $N$ is defined by
\begin{equation}
  bN=Q(1-g)-\sum_i\alpha_i
\end{equation}
 for worldsheets with $g$ handles.
 Although the expression (\ref{GLI}) can be used to evaluate
 correlators by first assuming $N$ to be a non-negative integer
 and then extending the result to generic $N$, we do not use it
 this way.
 We would rather read off from it one important property 
 that the correlator diverges when a non-negative integer insertions
 of $S_{\rm int}$ can screen the non-conserving Liouville momentum,
 and the residue of the divergence is given by the free field
 correlator with an appropriate number of $S_{\rm int}$ inserted.

\subsection{Two-point functions on a sphere}

   Here we re-derive the basic correlation functions on a sphere
 which were obtained by \cite{Poghosian,RS} as an introduction
 to our method to analyze the theory on a disc.
 The most important among them are the two-point functions.
 We first consider the following one:
\begin{equation}
  \vev{V_{\alpha_1}(z_1)V_{\alpha_2}(z_2)}
 = |z_{12}|^{-4h_{\alpha_1}}2\pi
   \left\{\delta(p_1+p_2) +\delta(p_1-p_2)D(\alpha_1)   \right\}
  ~~~~(\alpha_i=\tfrac{Q}{2}+ip_i).
\end{equation}
 The global superconformal symmetry yields some
 relations between two-point correlators:
\begin{eqnarray}
  \alpha_1\alpha_2\vev{\psi V_{\alpha_1}(z_1)\psi V_{\alpha_2}(z_2)}
 &=& 2h_{\alpha_1}z_{12}^{-1}
    \vev{V_{\alpha_1}(z_1)V_{\alpha_2}(z_2)}, \nonumber \\
  \alpha_1\alpha_2\vev{\bar{\psi}V_{\alpha_1}(z_1)\bar{\psi}V_{\alpha_2}(z_2)}
 &=& 2h_{\alpha_1}\bar{z}_{12}^{-1}
    \vev{V_{\alpha_1}(z_1)V_{\alpha_2}(z_2)}, \\
  \alpha_1^2\alpha_2^2
  \vev{\psi\bar{\psi}V_{\alpha_1}(z_1)\psi\bar{\psi}V_{\alpha_2}(z_2)}
 &=& -4h_{\alpha_1}^2|z_{12}|^{-2}
    \vev{V_{\alpha_1}(z_1)V_{\alpha_2}(z_2)}, \nonumber
\end{eqnarray}
 so that all the two-point functions of NSNS sector vertices
 are described by a single structure constant, $D(\alpha)$.
 To study the two-point functions of RR-sector vertices, we adopt
 the following convention.
 We first define the spin fields $\bar{\sigma}^\pm$ in the
 right-moving sector by the equations
\begin{equation}
  \bar{\sigma}^\pm(0)\bar{\psi}(z)\sim
  \frac{i\bar{\sigma}^\mp(0)}{\sqrt{2}\bar{z}^{1/2}}.
\label{sxpb}
\end{equation}
 Then the spin fields,
\begin{equation}
  \Theta_\alpha^{\ep\bep}(z,\bar{z})\equiv
  \sigma^{\ep\bep}e^{\alpha\phi}(z,\bar{z})\equiv
  \sigma^\ep\bar{\sigma}^{\bep}e^{\alpha\phi}(z,\bar{z}),~~~
  (\ep,\bep=\pm)
\end{equation}
 are shown to satisfy the OPE relations
\begin{equation}
\begin{array}{rcl}
  T_F(z)\Theta_\alpha^{\ep,\bep}(0)
 &\sim& \mfrac{p_\alpha\Theta_\alpha^{-\ep,\bep}(0)}{\sqrt{2}z^{3/2}},\\
  -i\Theta_\alpha^{\ep\bep}(0)\bar{T}_F(\bar{z})
 &\sim& \mfrac{p_\alpha\Theta_\alpha^{\ep,-\bep}(0)}{\sqrt{2}\bar{z}^{3/2}}.
\end{array}
\end{equation}
 We assume that $\sigma^+,\bar{\sigma}^+$ commute and
 $\sigma^-,\bar{\sigma}^-$ anti-commute with fermions.
 Thus $\Theta_\alpha^{\pm\pm}$ commute with fermions while
 $\Theta_\alpha^{\pm\mp}$ anti-commute.
 Using this, the superconformal Ward identity becomes
\begin{eqnarray}
\lefteqn{
  \oint\frac{dw}{2\pi i}(w-z)^{\frac{1}{2}}(w-z')^{\frac{1}{2}}T_F(w)
  \Theta_{\alpha }^{\ep\bep}(z)
  \Theta_{\alpha'}^{\ep'\bep'}(z')\left\{\cdots\right\}
} \nonumber \\
 &=& \frac{1}{\sqrt{2}}(z-z')^{\frac{1}{2}}
    \left[p_{\alpha}
    \Theta_{\alpha}^{-\ep,\bep}(z)
    \Theta_{\alpha'}^{\ep'\bep'}(z')\left\{\cdots\right\}
  +i\ep\bep p_{\alpha'}
    \Theta_{\alpha}^{\ep\bep}(z)
    \Theta_{\alpha'}^{-\ep',\bep'}(z')\left\{\cdots\right\}\right]
   \nonumber \\
 & &+\ep\bep\ep'\bep'
    \Theta_{\alpha}^{\ep\bep}(z)
    \Theta_{\alpha'}^{\ep'\bep'}(z')
    \oint\frac{dw}{2\pi i}(w-z)^{\frac{1}{2}}(w-z')^{\frac{1}{2}}
     T_F(w)\left\{\cdots\right\},
   \nonumber \\
\lefteqn{
  \oint\frac{d\bar{w}}{2\pi i}(\bar{w}-\bar{z})^{\frac{1}{2}}
 (\bar{w}-\bar{z}')^{\frac{1}{2}}
  \left\{\cdots\right\}\Theta_{\alpha }^{\ep\bep}(z)
       \Theta_{\alpha'}^{\ep'\bep'}(z')\bar{T}_F(\bar{w})
} \nonumber \\
 &=& \frac{1}{\sqrt{2}}(\bar{z}-\bar{z}')^{\frac{1}{2}}
    \left[\left\{\cdots\right\} p_{\alpha'}
    \Theta_{\alpha}^{\ep\bep}(z)
    \Theta_{\alpha'}^{\ep',-\bep'}(z')
  + \left\{\cdots\right\}i\ep'\bep' p_{\alpha}
    \Theta_{\alpha}^{\ep,-\bep}(z)
    \Theta_{\alpha'}^{\ep',\bep'}(z')\right] \nonumber \\
 & &+\ep\bep\ep'\bep'
    \oint\frac{dw}{2\pi i}(w-z)^{\frac{1}{2}}(w-z')^{\frac{1}{2}}
    \left\{\cdots\right\}\bar{T}_F(\bar{w})\cdot
    \Theta_{\alpha}^{\ep\bep}(z)
    \Theta_{\alpha'}^{\ep'\bep'}(z').
\end{eqnarray}
 The normalization of spin fields is given by the following
 correlators:
\begin{equation}
  \vev{\sigma^{\pm\pm}(z)\sigma^{\pm\pm}(0)}_{\rm free}
 =|z|^{-\frac{1}{4}},~~~
  \vev{\sigma^{\pm\mp}(z)\sigma^{\pm\mp}(0)}_{\rm free}
 =i|z|^{-\frac{1}{4}},
\label{ss}
\end{equation}
 where one should be careful for that only Grassmann-even combinations
 can have non-vanishing correlators.
 Note also that all these are related via superconformal transformations.
 From this, we put the following ansatz for the two-point functions
 of RR vertices:
\begin{equation}
\begin{array}{rcl}
    -i\vev{\Theta_{\alpha_1}^{\pm\mp}(z_1)\Theta_{\alpha_2}^{\pm\mp}(z_2)}
 =  \vev{\Theta_{\alpha_1}^{\pm\pm}(z_1)\Theta_{\alpha_2}^{\pm\pm}(z_2)}
 &=& |z_{12}|^{-4h_{\alpha_1}-\frac{1}{4}}\cdot 2\pi \delta(p_1+p_2), \\
     i\vev{\Theta_{\alpha_1}^{\pm\mp}(z_1)\Theta_{\alpha_2}^{\mp\pm}(z_2)}
 =  \vev{\Theta_{\alpha_1}^{\pm\pm}(z_1)\Theta_{\alpha_2}^{\mp\mp}(z_2)}
 &=& |z_{12}|^{-4h_{\alpha_1}-\frac{1}{4}}\cdot
    2\pi\delta(p_1-p_2)\tilde{D}(\alpha_1).
\end{array}
\label{2pS}
\end{equation}
 These two-point functions give relations between operators
 carrying Liouville momentum $\alpha$ and $Q-\alpha$:
\begin{equation}
\begin{array}{rcl}
 D(\alpha)
&=&\mfrac{V_\alpha}{V_{Q-\alpha}}  
 = \mfrac{\alpha\psi V_\alpha}{(Q-\alpha)\psi V_{Q-\alpha}}  
 = \mfrac{\alpha\bar{\psi}V_\alpha}{(Q-\alpha)\bar{\psi}V_{Q-\alpha}}  
 = \mfrac{\alpha^2\psi\bar{\psi}V_\alpha}
        {(Q-\alpha)^2\psi\bar{\psi}V_{Q-\alpha}},\\
 \tilde{D}(\alpha)
&=&\mfrac{\ep\bep\Theta_\alpha^{\ep\bep}}{\Theta_{Q-\alpha}^{-\ep,-\bep}}.
\end{array}
\label{rfc}
\end{equation}
 They are referred to as the reflection relation in what follows.
 $D(\alpha), \tilde{D}(\alpha)$ are called the reflection coefficients.

   The ansatz (\ref{2pS}) for the two-point functions of spin fields
 might seem peculiar at first sight, because it leads to the reflection
 relation which flips the indices $\ep,\bep$ as well as the momentum.
 In the following we obtain the reflection coefficients
 using some properties of degenerate primary fields, and there
 we will convince ourselves that the above ansatz is the only one
 which is consistent with the OPEs involving degenerate operators.

\paragraph{Degenerate fields and their OPEs}
   Let us summarize here some basic properties of
 operators belonging to degenerate representations of
 superconformal algebra.
 As was found in \cite{BKT,FQS,Nam}, they are given by the following
 Liouville momentum $\alpha_{r,s}$
\begin{equation}
  \alpha_{r,s} \equiv \frac{1}{2}(Q-rb-sb^{-1}),
\end{equation}
 in close analogy with the case of bosonic Liouville theory.
 The difference is that the degenerate representations with odd
 $r+s$ sit in the R sector, while those with even $r+s$ are in the
 NS sector.
 They are known to have null states at level $rs/2$.
 The corresponding vertex operators are $V_{\alpha_{r,s}}~(r+s$ even)
 or $\Theta_{\alpha_{r,s}}^\pm~(r+s$ odd).

   We will frequently use the most fundamental degenerate operators
 $\Theta_{-b/2}^\pm$ (and $\Theta_{-1/2b}^\pm$)
 in the following analysis.
 The property of the corresponding degenerate primary
 {\it states} $\ket{2,1}_\pm$ are summarized as follows:
\begin{equation}
  G_0\ket{2,1}_\pm = \tfrac{i(2b^2+1)}{2\sqrt{2}b}\ket{2,1}_\mp,~~~~
  G_{-1}\ket{2,1}_\pm + \tfrac{i\sqrt{2}}{b}L_{-1}\ket{2,1}_\mp =0.
\end{equation}
 We first discuss the OPEs involving these degenerate operators
 to find out the expressions for reflection coefficients.
 Then in later sections we will give a detailed analysis of
 the four-point functions involving them by solving
 the associated differential equations.
  
   Here we just assume that the OPEs of $\Theta_{-b/2}^{\ep\bep}$
 with arbitrary primary fields or spin fields yield only two
 discrete terms.
 This assumption will be justified in later sections
 by analyzing the four-point functions.
 We begin with one particular example:
\begin{equation}
 \Theta_{-b/2}^{++}(z_1)V_\alpha(z_2)
 \sim |z_{12}|^{b\alpha}C_+(\alpha)\Theta_{\alpha-b/2}^{++}(z_2)
    + |z_{12}|^{b(Q-\alpha)}C_-(\alpha)\Theta_{\alpha+b/2}^{--}(z_2),
\end{equation}
 where the coefficients $C_\pm(\alpha)$ are calculable
 using the techniques of free CFT as proposed in \cite{T}
 and utilized in the analysis of bosonic Liouville theory
 in \cite{FZZ}:
\begin{eqnarray}
  C_+(\alpha) &=& \lim_{z_1\rightarrow z_2}
   \frac{\vev{\Theta_{Q-\alpha+b/2}^{++}(w)
              \Theta_{-b/2}^{++}(z_1)V_\alpha(z_2)}_{\rm free}}
        {|z_{12}|^{b\alpha}
         \vev{\Theta_{Q-\alpha+b/2}^{++}(w)
              \Theta_{\alpha-b/2}^{++}(z_2)}_{\rm free}} =1, \nonumber \\
  C_-(\alpha) &=& \lim_{z_1\rightarrow z_2}
   \frac{\vev{\Theta_{Q-\alpha-b/2}^{--}(w)
              \Theta_{-b/2}^{++}(z_1)
              V_\alpha(z_2)(-S_{\rm int})}_{\rm free}}
        {|z_{12}|^{b\alpha}
         \vev{\Theta_{Q-\alpha-b/2}^{--}(w)
              \Theta_{\alpha+b/2}^{--}(z_2)}_{\rm free}} \nonumber \\
  &=& \mu\pi b^2\gamma(\tfrac{bQ}{2})\gamma(1-b\alpha)
      \gamma(b\alpha-\tfrac{bQ}{2}).
\label{CNS}
\end{eqnarray}
 Here we introduced the notation $\gamma(x)\equiv \Gamma(x)/\Gamma(1-x)$.
 The free field correlator in the numerator can be evaluated
 by using
\begin{equation}
   \vev{\sigma^{\pm\pm}(z_1)\sigma^{\mp\mp}(z_2)\psi\bar{\psi}(z_3)}
 =i\vev{\sigma^{\pm\mp}(z_1)\sigma^{\mp\pm}(z_2)\psi\bar{\psi}(z_3)}
 =\frac{i}{2}|z_{12}|^{3/4}|z_{13}z_{23}|^{-1},
\end{equation}
 which follows from (\ref{pxs}), (\ref{sxpb}) and (\ref{ss}).
 Summarizing similar OPE relations we obtain
\begin{eqnarray}
 \Theta_{-b/2}^{\ep\bep}(z_1)V_\alpha(z_2)
 &\sim&  |z_{12}|^{b\alpha}\Theta_{\alpha-b/2}^{\ep\bep}(z_2)
       +\ep\bep|z_{12}|^{b(Q-\alpha)}C_-(\alpha)
        \Theta_{\alpha+b/2}^{-\ep,-\bep}(z_2),
\label{SxV}
\end{eqnarray}
 Taking the superconformal transformation of both sides we obtain
\begin{eqnarray}
\lefteqn{
 \sqrt{2}z_{21}^{1/2}\Theta_{-b/2}^{\ep\bep}(z_1)\alpha\psi V_\alpha(z_2)
}\nonumber \\
 &\sim& \ep\bep|z_{12}|^{b\alpha}\alpha
        \Theta_{\alpha-b/2}^{-\ep,\bep}(z_2)
    -   |z_{12}|^{b(Q-\alpha)}(Q-\alpha)C_-(\alpha)
        \Theta_{\alpha+b/2}^{\ep,-\bep}(z_2),
 \nonumber \\
\lefteqn{
 -i\sqrt{2}\bar{z}_{21}^{1/2}\Theta_{-b/2}^{\ep\bep}(z_1)
 \alpha\bar{\psi} V_\alpha(z_2)
}\nonumber \\
 &\sim&
     |z_{12}|^{b\alpha}\alpha
        \Theta_{\alpha-b/2}^{\ep,-\bep}(z_2)
  -\ep\bep |z_{12}|^{b(Q-\alpha)}(Q-\alpha)C_-(\alpha)
        \Theta_{\alpha+b/2}^{-\ep,\bep}(z_2),
 \nonumber \\
\lefteqn{
 -2i|z_{21}|\Theta_{-b/2}^{\ep\bep}(z_1)\alpha^2\psi\bar{\psi}V_\alpha(z_2)
}\nonumber \\
 &\sim& \ep\bep|z_{12}|^{b\alpha}
        \alpha^2\Theta_{\alpha-b/2}^{-\ep,-\bep}(z_2)
       +|z_{12}|^{b(Q-\alpha)}(Q-\alpha)^2
        C_-(\alpha)\Theta_{\alpha+b/2}^{\ep\bep}(z_2).
\end{eqnarray}
 Combining them with the reflection relations (\ref{rfc}),
 we obtain the following recursion relations
 between structure constants:
\begin{equation}
  D(\alpha)=C_-(\alpha)\tilde{D}(\alpha+b/2),~~~
  \tilde{D}(\alpha-b/2)=C_-(Q-\alpha)D(\alpha).
\label{rcD}
\end{equation}

   To find the OPE of $\Theta_{-b/2}^{\ep\bep}$ with
 generic spin fields, let us start with
\begin{equation}
 -2\Theta_{-b/2}^{+-}(z_1)\Theta_\alpha^{-+}(z_2)
 \sim \tilde{C}_+(\alpha)|z_{12}|^{b\alpha+\frac{3}{4}}
      \psi\bar{\psi}V_{\alpha-b/2}(z_2)
     +\tilde{C}_-(\alpha)|z_{12}|^{b(Q-\alpha)-\frac{1}{4}}
      V_{\alpha+b/2}(z_2).
\end{equation}
 The free field integrals with screenings give
\begin{equation}
 \tilde{C}_+(\alpha)=1,~~~
 \tilde{C}_-(\alpha)=2i\mu\pi b^2\gamma(\tfrac{bQ}{2})
  \gamma(\tfrac{1}{2}-b\alpha)\gamma(b\alpha-\tfrac{b^2}{2})
  = 2iC_-(Q-\alpha-b/2).
\end{equation}
 Collecting similar OPE relations we have
\begin{eqnarray}
 -2\Theta_{-b/2}^{\pm\mp}(z_1)\Theta_\alpha^{\mp\pm}(z_2)
 &\sim&
 2i\Theta_{-b/2}^{\pm\pm}(z_1)\Theta_\alpha^{\mp\mp}(z_2)\nonumber \\
 &\sim&
   |z_{12}|^{b\alpha+\frac{3}{4}}\psi\bar{\psi}V_{\alpha-b/2}(z_2)
 +\tilde{C}_-(\alpha)|z_{12}|^{b(Q-\alpha)-\frac{1}{4}}V_{\alpha+b/2}(z_2).
\label{SxS}
\end{eqnarray}
 Taking its superconformal transformations we obtain
\begin{eqnarray}
\lefteqn{
 -\sqrt{2}\Theta_{-b/2}^{\pm\pm}(z_1)\Theta_\alpha^{\pm\mp}(z_2)
 ~\sim~
 i\sqrt{2}\Theta_{-b/2}^{\pm\mp}(z_1)\Theta_\alpha^{\pm\pm}(z_2)
} \hskip15mm \nonumber \\
 &\sim&
   \bar{z}_{12}^{\frac{1}{2}}|z_{12}|^{b\alpha-\frac{1}{4}}
  \bar{\psi}V_{\alpha-b/2}(z_2)
 +\frac{(2\alpha+b)\tilde{C}_-(\alpha)}{2(2Q-2\alpha-b)}
   z_{12}^{\frac{1}{2}}|z_{12}|^{b(Q-\alpha)-\frac{1}{4}}
  \psi V_{\alpha+b/2}(z_2),
 \nonumber \\
\lefteqn{
 -\sqrt{2}\Theta_{-b/2}^{\mp\pm}(z_1)\Theta_\alpha^{\pm\pm}(z_2)
 ~\sim~
 i\sqrt{2}\Theta_{-b/2}^{\pm\pm}(z_1)\Theta_\alpha^{\mp\pm}(z_2)
} \hskip15mm \nonumber \\
 &\sim&
 - z_{12}^{\frac{1}{2}}|z_{12}|^{b\alpha-\frac{1}{4}}
  \psi V_{\alpha-b/2}(z_2)
 +\frac{(2\alpha+b)\tilde{C}_-(\alpha)}{2(2Q-2\alpha-b)}
   \bar{z}_{12}^{\frac{1}{2}}|z_{12}|^{b(Q-\alpha)-\frac{1}{4}}
  \bar{\psi} V_{\alpha+b/2}(z_2),
 \nonumber \\
\lefteqn{
 -\Theta_{-b/2}^{\pm\pm}(z_1)\Theta_\alpha^{\pm\pm}(z_2)
 ~\sim~
 i\Theta_{-b/2}^{\pm\mp}(z_1)\Theta_\alpha^{\pm\mp}(z_2)
} \hskip15mm \nonumber \\
 &\sim&
 - |z_{12}|^{b\alpha-\frac{1}{4}}V_{\alpha-b/2}(z_2)
 +\frac{(2\alpha+b)^2\tilde{C}_-(\alpha)}{4(2Q-2\alpha-b)^2}
   |z_{12}|^{b(Q-\alpha)+\frac{3}{4}}
  \psi\bar{\psi} V_{\alpha+b/2}(z_2).
\end{eqnarray}
 Combining them with the reflection relations(\ref{rfc}), we obtain
 the same set of recursion relations as (\ref{rcD}).
 Here one can also see that the reflection of the Liouville momentum
 of spin fields should be accompanied by the flip of signs $\ep,\bep$,
 because in the right hand side of the above OPEs there are
 always two NS operators of opposite fermion number
 (when counted for either one of the left/right sectors separately).

   A simple solution of (\ref{rcD}) satisfying the unitarity
\begin{equation}
  D(\alpha)D(Q-\alpha)=\tilde{D}(\alpha)\tilde{D}(Q-\alpha)=1
\end{equation}
 and exhibiting the $b\leftrightarrow 1/b$ duality
 can be found rather easily:
\begin{eqnarray}
  D(\alpha) &=& 
 -\left(\mu\pi\gamma(bQ/2)\right)^{\frac{Q-2\alpha}{b}}
  \frac{\Gamma(b(\alpha-\frac{Q}{2}))
        \Gamma(\frac{1}{b}(\alpha-\frac{Q}{2}))}
       {\Gamma(-b(\alpha-\frac{Q}{2}))
        \Gamma(-\frac{1}{b}(\alpha-\frac{Q}{2}))}, \\
  \tilde{D}(\alpha) &=&
  \left(\mu\pi\gamma(bQ/2)\right)^{\frac{Q-2\alpha}{b}}
  \frac{\Gamma(\frac{1}{2}+b(\alpha-\frac{Q}{2}))
        \Gamma(\frac{1}{2}+\frac{1}{b}(\alpha-\frac{Q}{2}))}
       {\Gamma(\frac{1}{2}-b(\alpha-\frac{Q}{2}))
        \Gamma(\frac{1}{2}-\frac{1}{b}(\alpha-\frac{Q}{2}))}.
\end{eqnarray}

   Before moving on to the analysis of three-point functions,
 we note that the correlators involving NS-R vertices are difficult
 to analyze since the screening operator becomes double-valued
 around them.
 Therefore, they are not well-defined vertex operators
 from the viewpoint of a perturbed free CFT.

   We also note that the following relations for products
 of two spin fields hold:
\begin{equation}
  \Theta_{\alpha_1}^{\ep\bep}(z_1)
  \Theta_{\alpha_2}^{\tau\btau}(z_2)
 =-i\bep\tau\Theta_{\alpha_1}^{-\ep,\bep}(z_1)
            \Theta_{\alpha_2}^{-\tau,\btau}(z_2)
 =-i\bep\tau\Theta_{\alpha_1}^{\ep,-\bep}(z_1)
            \Theta_{\alpha_2}^{\tau,-\btau}(z_2),
\label{SS}
\end{equation}
 in all the expressions for two-point functions
 and OPEs obtained so far.
 We assume this to hold in arbitrary correlation functions
 containing two spin fields.
 As an evidence, the analysis of four-point
 functions becomes much simpler without any contradiction
 if we employ this relation.
 However, one should not expect this relation to hold
 in correlators containing more than two spin fields.

\subsection{Three-point functions on a sphere}

   We first put the following ansatz for them:
\begin{equation}
\begin{array}{rcl}
  \vev{V_{\alpha_1}V_{\alpha_2}V_{\alpha_3}}
 &=& C_1(\alpha_1,\alpha_2,\alpha_3), \\
  -i\alpha_1\vev{\psi V_{\alpha_1}V_{\alpha_2}V_{\alpha_3}}
 &=& C_2(\alpha_1,\alpha_2,\alpha_3), \\
  -\alpha_1^2\vev{\psi\bar{\psi}V_{\alpha_1}V_{\alpha_2}V_{\alpha_3}}
 &=& C_3(\alpha_1,\alpha_2,\alpha_3),
\end{array}
~~~
\begin{array}{rcl}
  \vev{V_{\alpha_1}
       \Theta_{\alpha_2}^{\pm\pm}
       \Theta_{\alpha_3}^{\mp\mp}}
 &=& \tilde{C}_1(\alpha_1;\alpha_2,\alpha_3), \\
  \vev{V_{\alpha_1}
       \Theta_{\alpha_2}^{\pm\pm}
       \Theta_{\alpha_3}^{\pm\mp}}
 &=& \tilde{C}_2(\alpha_1;\alpha_2,\alpha_3), \\
  \vev{V_{\alpha_1}
       \Theta_{\alpha_2}^{\pm\pm}
       \Theta_{\alpha_3}^{\pm\pm}}
 &=& \tilde{C}_3(\alpha_1;\alpha_2,\alpha_3),
\end{array}
\end{equation}
 where we omit the coordinate dependence which can easily be restored
 knowing the conformal weights of operators:
\begin{equation}
  \vev{{\cal O}_1(z_1){\cal O}_2(z_2){\cal O}_3(z_3)}
 \sim z_{12}^{h_3-h_1-h_2}
      z_{23}^{h_1-h_2-h_3}
      z_{13}^{h_2-h_3-h_1}
      \bar{z}_{12}^{\bar{h}_3-\bar{h}_1-\bar{h}_2}
      \bar{z}_{23}^{\bar{h}_1-\bar{h}_2-\bar{h}_3}
      \bar{z}_{13}^{\bar{h}_2-\bar{h}_3-\bar{h}_1}
\end{equation}
 Other three-point functions are related to the above expressions
 via left-right symmetry, or are obtained by taking the
 superconformal transformations of them.
 Using the superconformal symmetry we can also find that
 $C_{1,2,3}(\alpha_i)$ are symmetric in the three arguments.

   To obtain them we again use the degenerate field
 $\Theta^{\ep\bep}_{-b/2}$.
 We focus on the left-moving sector of the theory and consider
 the four-point functions in chiral CFT:
\begin{equation}
\begin{array}{rcl}
  \vev{V_{\alpha_3}(z_3)V_{\alpha_2}(z_2)
       \Theta_{-b/2}^{\ep}(z_0)\Theta_{\alpha_1}^{\tau}(z_1)}
 &=& f_{00}^{\ep\tau}(z_i), \\
  -i\alpha_2\vev{V_{\alpha_3}(z_3)\psi V_{\alpha_2}(z_2)
       \Theta_{-b/2}^{\ep}(z_0)\Theta_{\alpha_1}^{\tau}(z_1)}
 &=& f_{01}^{\ep\tau}(z_i), \\
  -i\alpha_3\vev{\psi V_{\alpha_3}(z_3)V_{\alpha_2}(z_2)
       \Theta_{-b/2}^{\ep}(z_0)\Theta_{\alpha_1}^{\tau}(z_1)}
 &=& f_{10}^{\ep\tau}(z_i), \\
  -\alpha_2\alpha_3\vev{\psi V_{\alpha_3}(z_3)\psi V_{\alpha_2}(z_2)
       \Theta_{-b/2}^{\ep}(z_0)\Theta_{\alpha_1}^{\tau}(z_1)}
 &=& f_{11}^{\ep\tau}(z_i).
\end{array}
\end{equation}
 As was shown in \cite{Poghosian} and reviewed in the following,
 a set of differential equations among them can be derived from
 the superconformal symmetry and the degeneracy of $\Theta_{-b/2}$
 which has two independent solutions.
 This justifies the assumption in the previous paragraph that the OPEs
 involving $\Theta_{-b/2}^{\ep\bep}$ yield only two discrete terms.
 However, in order to obtain the three-point structure constants
 it is sufficient to know that the first four of the above sixteen,
 $f_{00}^{\ep\tau}$, satisfies an ordinary hypergeometric
 differential equation.
 If we simply {\it assume} this we can skip the lengthy
 analysis of the differential equation and easily write down
 the solution, since the behavior when $z_0$ approach either
 of $z_{1,2,3}$ is known from the OPE formula.

\paragraph{Differential equation for correlators $\vev{VV\Theta\Theta}$}
   Firstly, the above sixteen functions are related
 via global superconformal transformation.
 By multiplying
 $\displaystyle\oint\mfrac{dw}{2\pi i}(w-z_0)^{1/2}(w-z_1)^{1/2}T_F(w)$
 onto the product of operators in the bracket we find
\begin{eqnarray}
  z_{30}^{1/2}z_{31}^{1/2}f_{10}^{\ep\tau}
 +z_{20}^{1/2}z_{21}^{1/2}f_{01}^{\ep\tau}
 +\frac{z_{01}^{1/2}}{\sqrt{2}}
 (p_{-b/2}-\ep\tau p_{\alpha_1})f_{00}^{-\ep,\tau}
 &=& 0, \nonumber \\
  z_{30}^{1/2}z_{31}^{1/2}f_{11}^{\ep\tau}
 +z_{20}^{1/2}z_{21}^{1/2}
 (\partial_2+\frac{h_2}{z_{20}}+\frac{h_2}{z_{21}})f_{00}^{\ep\tau}
 -\frac{z_{01}^{1/2}}{\sqrt{2}}
 (p_{-b/2}-\ep\tau p_{\alpha_1})f_{01}^{-\ep,\tau}
 &=& 0, \nonumber \\
  z_{30}^{1/2}z_{31}^{1/2}
 (\partial_3+\frac{h_3}{z_{30}}+\frac{h_3}{z_{31}})f_{00}^{\ep\tau}
 -z_{20}^{1/2}z_{21}^{1/2}f_{11}^{\ep\tau}
 -\frac{z_{01}^{1/2}}{\sqrt{2}}
 (p_{-b/2}-\ep\tau p_{\alpha_1})f_{10}^{-\ep,\tau}
 &=& 0, \\
  z_{30}^{1/2}z_{31}^{1/2}
 (\partial_3+\frac{h_3}{z_{30}}+\frac{h_3}{z_{31}})f_{01}^{\ep\tau}
 -z_{20}^{1/2}z_{21}^{1/2}
 (\partial_2+\frac{h_2}{z_{20}}+\frac{h_2}{z_{21}})f_{10}^{\ep\tau}
 \nonumber \\ 
 +\frac{z_{01}^{1/2}}{\sqrt{2}}
 (p_{-b/2}-\ep\tau p_{\alpha_1})f_{11}^{-\ep,\tau}
 &=& 0.
\nonumber
\end{eqnarray}
 Here the relation (\ref{SS}) was assumed.
 By introducing a new set of functions $F_{ij}^{\ep\tau}(\eta)$ of
 the cross ratio $\eta=\frac{z_{01}z_{23}}{z_{03}z_{21}}$:
\begin{equation}
\begin{array}{rcr}
  f_{00}^{\ep\tau}
  &=& \prod_{i>j}z_{ij}^{\mu_{ij}}F_{00}^{\ep\tau},\\
  f_{01}^{\ep\tau}
  &=& \left(\frac{z_{30}^2z_{21}}
                 {z_{32}^2z_{01}z_{20}}\right)^{1/2}
    \prod_{i>j}z_{ij}^{\mu_{ij}}F_{01}^{\ep\tau},\\
  f_{10}^{\ep\tau}
  &=& \left(\frac{z_{30}z_{21}^2}{z_{01}z_{31}z_{32}^2}\right)^{1/2}
    \prod_{i>j}z_{ij}^{\mu_{ij}}F_{10}^{\ep\tau},\\
  f_{11}^{\ep\tau}
  &=& \left(\frac{z_{30}z_{21}}{z_{20}z_{31}z_{32}^2}\right)^{1/2}
      \prod_{i>j}z_{ij}^{\mu_{ij}}F_{11}^{\ep\tau},
\end{array}
~~~~
\begin{array}{rcl}
  \sum_{j\ne 0}\mu_{0j} &=& -2h_0-1/8,\\
  \sum_{j\ne 1}\mu_{1j} &=& -2h_1-1/8,\\
  \sum_{j\ne 2}\mu_{2j} &=& -2h_2,\\
  \sum_{j\ne 3}\mu_{3j} &=& -2h_3,
\end{array}
\end{equation}
 The above relations can be rewritten in a simpler way:
\begin{equation}
\begin{array}{rcl}
  \eta(1-\eta)\partial_\eta \equiv D \\
  F_{10}^{\ep\tau}+F_{01}^{\ep\tau}
 +\frac{1}{\sqrt{2}}(p_{-b/2}-\ep\tau p_{\alpha_1})\eta F_{00}^{-\ep,\tau}
 &=& 0, \\
  F_{11}^{\ep\tau}
 -\left\{D+\mu_{23}+\eta(h_2+\mu_{21})\right\}
  F_{00}^{\ep\tau}
 -\frac{1}{\sqrt{2}}(p_{-b/2}-\ep\tau p_{\alpha_1})F_{01}^{-\ep,\tau}
 &=& 0, \\
  \left\{D+\mu_{23}+\eta(h_3+\mu_{03})\right\}
   F_{00}^{\ep\tau}
 -F_{11}^{\ep\tau}
 -\frac{1}{\sqrt{2}}(p_{-b/2}-\ep\tau p_{\alpha_1})F_{10}^{-\ep,\tau}
 &=& 0, \\
  \left\{D+\mu_{23}-1
         +\eta(h_3+\mu_{03}+1)\right\}F_{01}^{\ep\tau}
 && \\
 +\left\{D+\mu_{23}-1+\eta(h_2+\mu_{12}+1)\right\}
   F_{10}^{\ep\tau}
 +\frac{1}{\sqrt{2}}(p_{-b/2}-\ep\tau p_{\alpha_1})
  \eta F_{11}^{-\ep,\tau}
 &=& 0.
\end{array}
\label{SCW}
\end{equation}

   Secondly, we derive another set of differential equations
 by translating the equation for null states
\begin{equation}
  (L_{-1}G_0+\frac{2b^2+1}{4}G_{-1})\ket{2,1}
 =(L_{-1}G_0+\frac{bp_{-b/2}}{2i}G_{-1})\ket{2,1}
 =0
\label{ns}
\end{equation}
 into a differential equation for correlators involving
 the corresponding degenerate operator.
 Here we deal with correlators
 involving only two spin fields one of which is degenerate,
 and use the following translation law:
\begin{equation}
\begin{array}{rcl}
  ({\cal L}_n\Theta_{-b/2}^{\ep})(z_0)\Theta_{\alpha_1}^{\tau}(z_1)
  &=& \displaystyle\oint_{z_0}\frac{dz_2}{2\pi i}
       z_{01}^{-1}z_{20}^{n+1}z_{21}^{-n+1}T(z_2)
     \Theta_{-b/2}^{\ep}(z_0)\Theta_{\alpha_1}^{\tau}(z_1), \\
  ({\cal G}_n\Theta_{-b/2}^{\ep})(z_0)\Theta_{\alpha_1}^{\tau}(z_1)
  &=& \displaystyle\oint_{z_0}\frac{dz_2}{2\pi i}
       z_{01}^{-1/2}z_{20}^{n+1/2}z_{21}^{-n+1/2}T_F(z_2)
     \Theta_{-b/2}^{\ep}(z_0)\Theta_{\alpha_1}^{\tau}(z_1),
\end{array}
\end{equation}
 and write down the differential equation in the following way:
\begin{equation}
 \vev{\cdots (\sqrt{2}ib^{-1}{\cal L}_{-1}\Theta_{-b/2}^{-\ep}
             +{\cal G}_{-1}\Theta_{-b/2}^{\ep})
      (z_0)\cdot\Theta_{\alpha_1}^{\tau}(z_1)}=0.
\end{equation}
 This yields a set of differential equations for the functions
 $F_{ij}^{\ep\tau}$:
\begin{eqnarray}
 0 &=&
 -i\sqrt{2}b^{-1}(D-\mu_{02}-(1-\eta)\mu_{03})F_{00}^{-\ep\tau}
  + \mfrac{1-\eta}{\eta}F_{01}^{\ep\tau}+\mfrac{1}{\eta}F_{10}^{\ep\tau},
 \nonumber \\
 0 &=&
 -i\sqrt{2}b^{-1}(D-\mu_{02}+\frac{1}{2}-(1-\eta)(\mu_{03}+1))F_{01}^{-\ep\tau}
 \nonumber \\&&~~~
 -(1-\eta)F_{11}^{\ep\tau}
 +(D+\mu_{23}+\eta(\mu_{12}+3h_2))F_{00}^{\ep\tau},
 \nonumber  \\
 0 &=&
 -i\sqrt{2}b^{-1}(D-\mu_{02}-(1-\eta)(\mu_{03}+\frac{1}{2}))F_{10}^{-\ep\tau}
\label{nsD} \\&&~~~
 -(1-\eta)(D+\mu_{23}+\eta(\mu_{03}-h_3))F_{00}^{\ep\tau}+F_{11}^{\ep\tau},
 \nonumber \\
 0 &=&
 -i\sqrt{2}b^{-1}(D-\mu_{02}+\frac{1}{2}-(1-\eta)(\mu_{03}+\frac{1}{2}))
          F_{11}^{-\ep\tau}
 \nonumber \\&&~~~
 +\frac{1-\eta}{\eta}(D+\mu_{23}-1+\eta(\mu_{03}+1-h_3))F_{01}^{\ep\tau}
 \nonumber \\&&~~~
 +\frac{1}{\eta}(D+\mu_{23}-1+\eta(\mu_{12}+1+3h_2))F_{10}^{\ep\tau}.
 \nonumber
\end{eqnarray}

   Obviously, (\ref{SCW}) and (\ref{nsD}) are redundant.
 Indeed, (\ref{SCW}) yields only two relations between
 $\{F_{00}^{\ep\tau},F_{01}^{-\ep\tau},
    F_{10}^{-\ep\tau},F_{11}^{\ep\tau}\}$, so that we are left with
 two independent functions out of four after imposing
 the superconformal symmetry.
 (\ref{nsD}) also contains only two independent equations,
 giving a second-ordered differential equation for
 one of the remaining two independent functions.
 Thus the solution for $f_{00}^{\ep\tau}$
 is written in terms of the hypergeometric function,
\begin{eqnarray}
  f_{00}^{\ep\tau}(z_{0,1,2,3})
&=&z_{12}^{h_3-h_1-h_2-h_0-1/8}
   z_{23}^{h_1-h_2-h_3+h_0+1/8}
   z_{13}^{h_2-h_3-h_1+h_0}
   z_{03}^{-2h_0-1/8}
 \nonumber \\ && \times
   \eta^{\mu_{01}}(1-\eta)^{\mu_{02}}F(A,B;C;\eta),
 \nonumber \\
  A &=& \mu_{01}+\mu_{02}+\frac{b\alpha_3}{2}+2h_0+\frac{1}{8},
 \nonumber \\
  B &=& \mu_{01}+\mu_{02}+\frac{b(Q-\alpha_3)}{2}+2h_0+\frac{1}{8},
 \nonumber \\
  C &=& bQ+2\mu_{01}+4h_{0}+\frac{1}{4},
 \nonumber \\
  \mu_{01} &=& \frac{bQ+\ep\tau b(2\alpha_1-Q)}{4}-\frac{1}{8} ~~~{\rm or}~~~
               \frac{bQ-\ep\tau b(2\alpha_1-Q)}{4}+\frac{3}{8},
 \nonumber \\
  \mu_{02} &=& \frac{b\alpha_2}{2} ~~~{\rm or}~~~ \frac{b(Q-\alpha_2)}{2}.
\end{eqnarray}
 The two choices for $\mu_{01}$ give two independent solutions,
 while two choices of $\mu_{02}$ lead to the same solution owing
 to the formula of hypergeometric functions.

~

   By taking its square in a crossing symmetric way we can construct
 the four-point function on a sphere involving a degenerate field,
\begin{eqnarray}
\lefteqn{
  \vev{V_{\alpha_3}(z_3)V_{\alpha_2}(z_2)
       \Theta_{-b/2}^{++}(z_0)\Theta_{\alpha_1}^{--}(z_1)} } \nonumber \\
 &=& |z_{12}|^{2(h_3-h_1-h_2-h_0-1/8)}
   |z_{23}|^{2(h_1-h_2-h_3+h_0+1/8)}
   |z_{13}|^{2(h_2-h_3-h_1+h_0)}
   |z_{03}|^{-4h_0-1/4}
 \nonumber \\ && \times
   \left\{P_1G_1(\alpha_1,\alpha_2,\alpha_3;\eta)
             G_1(\alpha_1,\alpha_2,\alpha_3;\bar{\eta})
  \right. \nonumber \\ && ~~~\left.
         +P_2G_2(Q-\alpha_1,\alpha_2,\alpha_3;\eta)
             G_2(Q-\alpha_1,\alpha_2,\alpha_3;\bar{\eta})
   \right\},
\label{VVSS}
\end{eqnarray}
 where $G_{1,2}$ are given by
 (we use the notation like $p_{-1-2+3}\equiv -p_1-p_2+p_3$)
\begin{equation}
\begin{array}{rcl}
 G_1(\alpha_i;\eta) &=&
 \eta^{\frac{b\alpha_1}{2}+\frac{3}{8}}
 (1-\eta)^{\frac{b\alpha_2}{2}}
 F(\frac{ib}{2}p_{+1+2+3}+\frac{3}{4},\,
   \frac{ib}{2}p_{+1+2-3}+\frac{3}{4};\,
   ibp_1+\frac{3}{2};\,\eta) \\
 G_2(\alpha_i;\eta) &=&
 \eta^{\frac{b\alpha_1}{2}-\frac{1}{8}}
 (1-\eta)^{\frac{b\alpha_2}{2}}
 F(\frac{ib}{2}p_{+1+2+3}+\frac{1}{4},\,
   \frac{ib}{2}p_{+1+2-3}+\frac{1}{4};\,
   ibp_1+\frac{1}{2}\,;\eta)
\end{array}
\end{equation}
 and the coefficients $P_{1,2}$ can be fixed from the crossing symmetry
 up to an overall normalization:
\begin{eqnarray}
 \frac{P_1}{P_2}&=&
 -(\tfrac{1}{2}+ibp_1)^{-2}
  \gamma(\tfrac{1}{2}-ibp_1)^2
 \nonumber \\ && \times
  \gamma(\tfrac{3}{4}+\tfrac{ib}{2}p_{+1+2+3})
  \gamma(\tfrac{3}{4}+\tfrac{ib}{2}p_{+1+2-3})
  \gamma(\tfrac{3}{4}+\tfrac{ib}{2}p_{+1-2+3})
  \gamma(\tfrac{3}{4}+\tfrac{ib}{2}p_{+1-2-3}).
\end{eqnarray}
 Considering in the same way, we find that there is no
 crossing symmetric solution for four point functions like
 $\vev{V_{\alpha_3}V_{\alpha_2}\Theta_{-b/2}^{++}\Theta_{\alpha_1}^{+-}}$,
 so that they must vanish.
 The correlator
 $\vev{V_{\alpha_3}V_{\alpha_2}\Theta_{-b/2}^{++}\Theta_{\alpha_1}^{++}}$
 are obtained simply by replacing $\alpha_1$ with $Q-\alpha_1$
 in the above.

   From the above solutions we can derive a recursion relation for
 three-point structure constants $C_i(\alpha_{1,2,3})$ and
 $\tilde{C}_i(\alpha_{1,2,3})$.
 From the fact that some four-point functions vanish it follows that
\begin{equation}
  C_2(\alpha_1,\alpha_2,\alpha_3)
 =\tilde{C}_2(\alpha_1,\alpha_2,\alpha_3)=0.
\end{equation}
 By comparing the limit $z_0\rightarrow z_1$ of (\ref{VVSS})
 with the OPE formula (\ref{SxS}) we find
\begin{eqnarray}
  \frac{P_1}{P_2}(p_1,p_2,p_3) &=&
 -\frac{1}{(\alpha_1-\frac{b}{2})^2}
  \frac{\tilde{C}_+(\alpha_1)C_3(\alpha_1-\frac{b}{2},\alpha_2,\alpha_3)}
       {\tilde{C}_-(\alpha_1)C_1(\alpha_1+\frac{b}{2},\alpha_2,\alpha_3)},
 \nonumber \\
  \frac{P_1}{P_2}(-p_1,p_2,p_3) &=&
  \frac{1}{(2Q-2\alpha_1-b)^2}
  \frac{\tilde{C}_-(\alpha_1)C_3(\alpha_1+\frac{b}{2},\alpha_2,\alpha_3)}
       {\tilde{C}_+(\alpha_1)C_1(\alpha_1-\frac{b}{2},\alpha_2,\alpha_3)}.
\end{eqnarray}
 These give a set of recursion relations for
 $C_1(\alpha_1,\alpha_2,\alpha_3)$ and
 $C_3(\alpha_1,\alpha_2,\alpha_3)$, whose solution can be expressed
 in terms of $\Up$ function introduced in \cite{DO,ZZ2}.
 It is defined by
\begin{eqnarray}
 \ln\Up(x)  &=& \int_0^\infty \frac{dt}{t}\left[
    e^{-2t}(\tfrac{Q}{2}-x)^2 -\frac{\sinh^2[(Q/2-x)t]}{\sinh[bt]\sinh[t/b]}
   \right],
\end{eqnarray}
 and satisfies the following relations
\begin{equation}
  \Up(x+b)=\Up(x)b^{1-2bx}\gamma(bx),~~~
  \Up(x+\tfrac{1}{b})=\Up(x)b^{2x/b-1}\gamma(x/b),~~~
  \Up(x) =\Up(Q-x).
\end{equation}
 It has zeroes at
\begin{equation}
  \Up(x)=0~~{\rm at}~~x=-mb-nb^{-1},~x=Q+mb+nb^{-1}~~(m,n\in {\bf Z}_{\ge 0}).
\end{equation}
 If we define
\begin{equation}
 \Up_\NS(x) = \Up(\tfrac{x}{2})\Up(\tfrac{x+Q}{2}),~~~
 \Up_\R(x)  = \Up(\tfrac{x+b}{2})\Up(\tfrac{x+b^{-1}}{2}),
\end{equation}
 the solution for the recursion relation can be expressed as
\begin{eqnarray}
  C_1(\alpha_i)
 &=&\left\{\mu\pi\gamma(\tfrac{bQ}{2})b^{1-b^2}\right\}^
     {\frac{Q-\Sigma\alpha_i}{b}}
    \hskip-4mm
    \frac{~~\Up_\NS'(0)}{\Up_\NS(\alpha_{1+2+3}-Q)}
 \frac{\Up_\NS(2\alpha_1)\Up_\NS(2\alpha_2)\Up_\NS(2\alpha_3)}
      {\Up_\NS(\alpha_{1+2-3})\Up_\NS(\alpha_{2+3-1})\Up_\NS(\alpha_{3+1-2})},
 \nonumber \\
  C_3(\alpha_i)
&=&i\left\{\mu\pi\gamma(\tfrac{bQ}{2})b^{1-b^2}\right\}^{\frac{Q-2A}{b}}
    \hskip-4mm
    \frac{~~2\Up_\NS'(0)}{\Up_\R(\alpha_{1+2+3}-Q)}
 \frac{\Up_\NS(2\alpha_1)\Up_\NS(2\alpha_2)\Up_\NS(2\alpha_3)}
      {\Up_\R(\alpha_{1+2-3})\Up_\R(\alpha_{2+3-1})\Up_\R(\alpha_{3+1-2})},
\label{VVV}
\end{eqnarray}
 where we used the notations like
 $\alpha_{1+2-3}\equiv \alpha_1+\alpha_2-\alpha_3$.
 The functions $\Up_\NS,\Up_\R$ are also useful to construct
 reflection-symmetric quantities because of the relations
\begin{eqnarray}
  D(\alpha) &=& (\mu\pi\gamma(\tfrac{bQ}{2})b^{1-b^2})^{\frac{Q-2\alpha}{b}}
                \frac{\Up_\NS(2\alpha)}{\Up_\NS(2Q-2\alpha)},
 \nonumber \\
  \tilde{D}(\alpha) &=&
  (\mu\pi\gamma(\tfrac{bQ}{2})b^{1-b^2})^{\frac{Q-2\alpha}{b}}
  \frac{\Up_\R(2\alpha)}{\Up_\R(2Q-2\alpha)}.
\end{eqnarray}

   The three-point structure constants containing spin fields
 can be obtained in a similar way.
 This time we take the limit $z_0\rightarrow z_2$ of 
 the solution (\ref{VVSS}).
 Using the formulae for hypergeometric functions we find
\begin{eqnarray}
\lefteqn{
  \vev{V_{\alpha_3}(z_3)V_{\alpha_2}(z_2)
       \Theta_{-b/2}^{++}(z_0)\Theta_{\alpha_1}^{--}(z_1)} } \nonumber \\
 &=& |z_{12}|^{2(h_3-h_1-h_2-h_0-1/8)}
   |z_{23}|^{2(h_1-h_2-h_3+h_0+1/8)}
   |z_{13}|^{2(h_2-h_3-h_1+h_0)}
   |z_{03}|^{-4h_0-1/4}
 \nonumber \\ && \hskip-5mm\times
   \left\{Q_1H(\alpha_1,\alpha_2,\alpha_3;1-\eta)
             H(\alpha_1,\alpha_2,\alpha_3;1-\bar{\eta})
   \right. \nonumber \\ && ~~~ \left.
         +Q_2H(\alpha_1,Q-\alpha_2,\alpha_3;1-\eta)
             H(\alpha_1,Q-\alpha_2,\alpha_3;1-\bar{\eta})
   \right\},
\end{eqnarray}
 where $H$ is given by
\begin{equation}
 H(\alpha_i;1-\eta) =
 \eta^{\frac{b\alpha_1}{2}+\frac{3}{8}}
 (1-\eta)^{\frac{b\alpha_2}{2}}
 F(\tfrac{ib}{2}p_{+1+2+3}+\tfrac{3}{4},\,
   \tfrac{ib}{2}p_{+1+2-3}+\tfrac{3}{4};\,
   ibp_2+1;\,1-\eta),
\end{equation}
 and the ratio $Q_1/Q_2$ reads
\begin{equation}
  \frac{Q_1}{Q_2} =
 b^2p_2^2\gamma(-ibp_2)^2
  \gamma(\tfrac{3}{4}+\tfrac{ib}{2}p_{1+2+3})
  \gamma(\tfrac{3}{4}+\tfrac{ib}{2}p_{1+2-3})
  \gamma(\tfrac{1}{4}+\tfrac{ib}{2}p_{-1+2+3})
  \gamma(\tfrac{1}{4}+\tfrac{ib}{2}p_{-1+2-3}).
\end{equation}
 By making a comparison with the OPE formula (\ref{SxV}) we find
 recursion relations
\begin{eqnarray}
  \frac{Q_1}{Q_2}(p_1,p_2,p_3) &=&
  \frac{C_+(\alpha_2)\tilde{C}_1(\alpha_3;\alpha_2-\frac{b}{2},\alpha_1)}
       {C_-(\alpha_2)\tilde{C}_3(\alpha_3;\alpha_2+\frac{b}{2},\alpha_1)},
 \nonumber \\
  \frac{Q_1}{Q_2}(-p_1,p_2,p_3) &=&
  \frac{C_+(\alpha_2)\tilde{C}_3(\alpha_3;\alpha_2-\frac{b}{2},\alpha_1)}
       {C_-(\alpha_2)\tilde{C}_1(\alpha_3;\alpha_2+\frac{b}{2},\alpha_1)}.
\end{eqnarray}
 This can be straightforwardly solved and we obtain
\begin{eqnarray}
  \tilde{C}_1(\alpha_3;\alpha_2,\alpha_1) =
  \left\{\mu\pi\gamma(\tfrac{bQ}{2})b^{1-b^2}\right\}^{\frac{Q-2A}{b}}
  \hskip-7mm
  \frac{\Up_\NS'(0)\Up_\NS(2\alpha_3)\Up_\R(2\alpha_2)\Up_\R(2\alpha_1)}
       {\Up_\R(\alpha_{1+2+3}-Q)\Up_\R(\alpha_{1+2-3})
        \Up_\NS(\alpha_{2+3-1})\Up_\NS(\alpha_{3+1-2})},
 \nonumber \\
  \tilde{C}_3(\alpha_3;\alpha_2,\alpha_1) =
  \left\{\mu\pi\gamma(\tfrac{bQ}{2})b^{1-b^2}\right\}^{\frac{Q-2A}{b}}
  \hskip-7mm
  \frac{\Up_\NS'(0)\Up_\NS(2\alpha_3)\Up_\R(2\alpha_2)\Up_\R(2\alpha_1)}
       {\Up_\NS(\alpha_{1+2+3}-Q)\Up_\NS(\alpha_{1+2-3})
        \Up_\R(\alpha_{2+3-1})\Up_\R(\alpha_{3+1-2})},
\label{VSS}
\end{eqnarray}
 We can also check that these three-point structure constants
 are consistent with the reflection symmetry.

   Although these three-point structure constants
 (\ref{VVV}) and (\ref{VSS}) have complicated form,
 each factor has a clear physical meaning.
 Note first that the zeroes of $\Up_\NS,\Up_\R$ are at
\begin{eqnarray}
  \Up_\NS(x)=0&&
  ~~{\rm at}~~ x=-mb-nb^{-1},~x=Q+mb+nb^{-1}~~(m+n~~{\rm even}), \nonumber \\
  \Up_\R(x)=0&&
  ~~{\rm at}~~ x=-mb-nb^{-1},~x=Q+mb+nb^{-1}~~(m+n~~{\rm odd}).
\end{eqnarray}
 Each three-point structure constant therefore has eight
 sequences of poles.
 This agree with our naive expectation, since by combining
 the perturbative consideration, $b\leftrightarrow b^{-1}$
 and the reflection symmetry we can easily guess that,
 for example, $C_1(\alpha_i)$ diverges at
\begin{equation}
 (\alpha_1 ~\mbox{or}~Q-\alpha_1)+
 (\alpha_2 ~\mbox{or}~Q-\alpha_2)+
 (\alpha_3 ~\mbox{or}~Q-\alpha_3) = Q-mb-nb^{-1}
 ~~~(m+n~~\mbox{even}).
\end{equation}
 The last condition is because of the fermionic nature of
 the screening operators.
 The pole structure of other structure constants can also be
 understood in the same way if we take into account that the
 reflection of spin fields also flips their Grassmann parity.

\paragraph{Differential equation for correlators
           $\vev{\Theta\Theta\Theta\Theta}$}
   Let us analyze here the correlation functions of
 four spin fields, one of which is degenerate.
 The reason for this is that we will use the solution
 to obtain the one-point structure constant on a disc
 in the next section.
 The solution can also be used to cross-check the three-point
 structure constants which were obtained previously.

   We first consider the holomorphic sector as in the previous case,
 and begin by introducing some notations:
\begin{equation}
  \vev{\Theta_{-b/2    }^{\ep_0}(z_0)
       \Theta_{\alpha_1}^{\ep_1}(z_1)
       \Theta_{\alpha_2}^{\ep_2}(z_2)
       \Theta_{\alpha_3}^{\ep_3}(z_3)}
 \equiv f^{\ep_0\ep_1\ep_2\ep_3}(z_i).
\label{SSSS}
\end{equation}
 Then we translate the expression for the null vector (\ref{ns})
 into differential equation for correlators.
 In doing this, note first that the Ramond algebra is generated by
 $L_1$ and $G_{-1}$.
 So it suffices to give the rule of translation for these two:
\begin{eqnarray}
  \vev{{\cal L}_1\Theta_{-b/2    }^{\ep_0}(z_0)\tprod_i
       \Theta_{\alpha_i}^{\ep_i}(z_i)}
 &=& \oint_{z_0}\frac{dz_4}{2\pi i}z_{40}^2
  \vev{T(z_4)
       \Theta_{-b/2    }^{\ep_0}(z_0)\tprod_i
       \Theta_{\alpha_i}^{\ep_i}(z_i)},
 \nonumber \\
  \vev{{\cal G}_{-1}\Theta_{-b/2    }^{\ep_0}(z_0)\tprod_i
       \Theta_{\alpha_i}^{\ep_i}(z_i)}
 &=& \oint_{z_0}\frac{dz_4}{2\pi i}
     \left(\frac{z_{41}z_{42}z_{43}}{z_{40}z_{01}z_{02}z_{03}}
     \right)^{\frac{1}{2}}\!\!\!
  \vev{T_F(z_4)
       \Theta_{-b/2    }^{\ep_0}(z_0)\tprod_i
       \Theta_{\alpha_i}^{\ep_i}(z_i)}.
\end{eqnarray}
 Using them, the rule for other generators can be obtained easily:
\begin{eqnarray}
\lefteqn{
  \vev{{\cal G}_0\Theta_{-b/2    }^{\ep_0}(z_0)\cdot\tprod_i
       \Theta_{\alpha_i}^{\ep_i}(z_i)} }\nonumber \\
 &=& \frac{1}{3}\oint_{z_0}\frac{dz_4}{2\pi i}
     \left(\frac{z_{40}z_{41}z_{42}z_{43}}{z_{01}z_{02}z_{03}}\right)^{1/2}
     \left(\frac{z_{01}}{z_{41}}
          +\frac{z_{02}}{z_{42}}
          +\frac{z_{03}}{z_{43}}\right)
  \vev{T_F(z_4)
       \Theta_{-b/2    }^{\ep_0}(z_0)\tprod_i
       \Theta_{\alpha_i}^{\ep_i}(z_i)},
 \nonumber \\
\lefteqn{
  \vev{{\cal L}_{-1}\Theta_{-b/2    }^{\ep_0}(z_0)\cdot\tprod_i
       \Theta_{\alpha_i}^{\ep_i}(z_i)} }\nonumber \\
 &=& \frac{1}{3}\oint_{z_0}\frac{dz_4}{2\pi i}
     \frac{z_{41}z_{42}z_{43}}{z_{01}z_{02}z_{03}}
     \left(\frac{z_{01}}{z_{41}}
          +\frac{z_{02}}{z_{42}}
          +\frac{z_{03}}{z_{43}}\right)
  \vev{T(z_4)
       \Theta_{-b/2    }^{\ep_0}(z_0)\tprod_i
       \Theta_{\alpha_i}^{\ep_i}(z_i)}.
\end{eqnarray}
 Then the equation for the null vector
 (where we use the notation $\mep\equiv-\ep$),
\begin{eqnarray}
  \sqrt{2}ib^{-1}{\cal L}_{-1}
  \vev{\Theta_{-b/2    }^{\mep_0}(z_0)\cdot\tprod_i
       \Theta_{\alpha_i}^{\ep_i}(z_i)}
  +{\cal G}_{-1}
  \vev{\Theta_{-b/2    }^{\ep_0}(z_0)\cdot\tprod_i
       \Theta_{\alpha_i}^{\ep_i}(z_i)}
 =0,
\end{eqnarray}
 can be recast into the form of a differential equation,
\begin{eqnarray}
&&
  \ep_0\left(\frac{z_{12}z_{13}}{z_{10}}\right)^{1/2}\hskip-3mm
  p_1f^{\ep_0\mep_1\ep_2\ep_3}
 +\ep_0\ep_1\left(\frac{z_{21}z_{23}}{z_{20}}\right)^{1/2}\hskip-3mm
  p_2f^{\ep_0\ep_1\mep_2\ep_3}
 +\ep_0\ep_1\ep_2\left(\frac{z_{31}z_{32}}{z_{30}}\right)^{1/2}\hskip-3mm
  p_3f^{\ep_0\ep_1\ep_2\mep_3}
 \nonumber \\
&& ~=~
  \frac{2i}{b}(z_{01}z_{02}z_{03})^{1/2}\left\{
  \partial_0
 +\frac{2h_0+\frac{1}{8}}{3}
  (z_{01}^{-1}+z_{02}^{-1}+z_{03}^{-1})\right\}
  f^{\mep_0\ep_1\ep_2\ep_3}.
\end{eqnarray}
 As in the previous case we rescale the functions
 $f^{\ep_0\ep_1\ep_2\ep_3}$ in the following way:
\begin{equation}
  f^{\ep_0\ep_1\ep_2\ep_3} = \prod_{i,j}z_{ij}^{\mu_{ij}}
  F^{\ep_0\ep_1\ep_2\ep_3}(\eta),~~~
  \eta = \frac{z_{01}z_{23}}{z_{03}z_{21}},~~~
  \sum_{i(\ne j)}\mu_{ij} = -2h_j-\frac{1}{8}.
\end{equation}
 Then the above equation can be rewritten into the form
\begin{eqnarray}
\lefteqn{ 2ib^{-1}\ep_0{\cal D}F^{\mep_0\ep_1\ep_2\ep_3} }\nonumber \\
&=&p_1(\eta-1)^{1/2}F^{\ep_0\mep_1\ep_2\ep_3}
 + \ep_1p_2(-\eta)^{1/2}F^{\ep_0\ep_1\mep_2\ep_3}
 + \ep_1\ep_2p_3\eta^{1/2}(1-\eta)^{1/2}F^{\ep_0\ep_1\ep_2\mep_3}
 ,\nonumber \\
 {\cal D} &=&
    \eta(1-\eta)\partial_\eta
   +(1-\eta)(\mu_{01}+\tfrac{1-2bQ}{8})
   -\eta(\mu_{02}+\tfrac{1-2bQ}{8}).
\label{nsFe}
\end{eqnarray}
 According to the signs of $\ep_i$, there are sixteen components
 of $F^{\ep_0\ep_1\ep_2\ep_3}$.
 The above equations separate them into two groups, each containing
 eight components with even(odd) number of minus signs in $\ep_i$.
 We try to reduce the number of independent components further
 by putting the assumption
\begin{equation}
  F^{\ep_0\ep_1\ep_2\ep_3}
 = c(\ep_0,\ep_1,\ep_2,\ep_3)F^{\mep_0\mep_1\mep_2\mep_3}.
\end{equation}
 The consistency with (\ref{nsFe}) yields
\begin{eqnarray}
&&  c(\ep_0,\ep_1,\ep_2,\ep_3)=c(\mep_0,\mep_1,\mep_2,\mep_3)=\pm 1,~~~
\nonumber \\&&
  c(\ep_0,\ep_1,\ep_2,\ep_3)
 =-c(\mep_0,\mep_1,\ep_2,\ep_3)
 =c(\mep_0,\ep_1,\mep_2,\ep_3)
 =-c(\mep_0,\ep_1,\ep_2,\mep_3).
\end{eqnarray}
 Denoting $c(+,+,+,+)=\xi$
 and $F_{0,1,2,3}=(F^{++++},F^{--++},F^{-+-+},F^{-++-})$ we have
\begin{equation}
\begin{array}{rrrrr}
  2ib^{-1}{\cal D}F_0&=&
 - ~~p_1(\eta-1)^{1/2}F_1&
 - ~~p_2(-\eta)^{1/2}F_2&
 - ~~p_3\eta^{1/2}(1-\eta)^{1/2}F_3,\\
  2ib^{-1}{\cal D}F_1&=&
 + ~~p_1(\eta-1)^{1/2}F_0&
 + \xi p_2(-\eta)^{1/2}F_3&
 - \xi p_3\eta^{1/2}(1-\eta)^{1/2}F_2,\\
  2ib^{-1}{\cal D}F_2&=&
 - \xi p_1(\eta-1)^{1/2}F_3&
 +~~ p_2(-\eta)^{1/2}F_0&
 + \xi p_3\eta^{1/2}(1-\eta)^{1/2}F_1,\\
  2ib^{-1}{\cal D}F_3&=&
 + \xi p_1(\eta-1)^{1/2}F_2&
 - \xi p_2(-\eta)^{1/2}F_1&
 + ~~p_3\eta^{1/2}(1-\eta)^{1/2}F_0.
\end{array}
\label{nsFi}
\end{equation}
 One can see that the above system of differential equations
 exhibits a symmetry in three generic spin fields.
 It is also consistent with the reflection symmetry: for example,
 $F_1$ obeys the same equation as that for $F_0$ with the signs
 of $p_2$ and $p_3$ flipped.
 Here we will not go into any further detail to determine
 $\xi$ or choose explicitly one appropriate
 branch for each square root, since different choices lead to
 different correlators and we would like to study their mutual relation
 later in detail.

   Let us step aside for a while and try another way to
 construct the correlators (\ref{SSSS}).
 Recall that the correlators are characterized by the analyticity
 and the asymptotic behavior around $z_0\sim z_{1,2,3}$ dictated
 by the OPE formula:
\begin{equation}
  f^{\ep_1\ep_2\ep_3\ep_4}(z_i) \stackrel{z_0\rightarrow z_i}{\sim}
  z_{0i}^{\frac{b\alpha_i}{2}-\frac{1}{8}},~~
  z_{0i}^{\frac{b\alpha_i}{2}+\frac{3}{8}},~~
  z_{0i}^{\frac{b(Q-\alpha_i)}{2}-\frac{1}{8}},~~
  z_{0i}^{\frac{b(Q-\alpha_i)}{2}+\frac{3}{8}}.
\label{asSxS}
\end{equation}
 In the previous analysis of reflection coefficients it was assumed
 that the OPE of $\Theta_{-b/2}^{\ep\bep}$ with generic spin fields
 (of definite chirality) yield only two discrete terms.
 However, since the differential equation is of the fourth order,
 there should be four independent solutions.
 Therefore, in solving the differential equation
 we should not adhere to the idea that each spin field
 in any correlator has a definite chirality.
 Thus we assume that the leading order behavior of the four independent
 solutions should be given by (\ref{asSxS}).

   Let us then put an assumption that the correlator can be expressed
 as a double contour integral of the following form:
\begin{equation}
   f^{\ep_1\ep_2\ep_3\ep_4}(z_i)=
  \prod_{i<j}z_{ij}^{\rho_{ij}}
  \int dwdw'\prod_i(z_i-w)^{\nu_i}(z_i-w')^{\nu'_i}(w-w')^\lambda
\end{equation}
 as one can guess by analogy with simpler cases where
 we encounter with hypergeometric functions.
 The global conformal invariance yields the conditions on
 the exponents $\nu_i,\nu'_i,\lambda$ and $\rho_{ij}$:
\begin{equation}
  \sum_i\nu_i=\sum_i\nu'_i=-\lambda-2,~~~~
 \sum_{j(\ne i)}\rho_{ij}+\nu_i+\nu'_i=-2h_i-\frac{1}{8}.
\end{equation}
 Analyzing the behavior at, say, $z_0\sim z_1$ we find that
 the double integral approximately breaks into several terms
 with different asymptotic behavior.
 In doing this, note first that the limit $z_0\rightarrow z_1$
 can also be viewed as the limit $z_2\rightarrow z_3$.
 Then there are two possibilities for $w$ to be
 either near $z_{0,1}$ or near $z_{2,3}$.
 Similarly, there are also two possibilities for $w'$,
 so that they altogether give four terms in the limit $z_0\sim z_1$:
\begin{equation}
 f^{\ep_1\ep_2\ep_3\ep_4}(z_i)\sim
 z_{01}^{\rho_{01}},~~
 z_{01}^{\rho_{01}+1+\nu'_0+\nu'_1},~~
 z_{01}^{\rho_{01}+1+\nu_0+\nu_1},~~
 z_{01}^{\rho_{01}+\lambda+2+\nu_0+\nu_1+\nu'_0+\nu'_1}.
\end{equation}
 Comparing this with the OPE formula, we can easily find that
 $\lambda$ must be $0$ or $\pm1$.
 Among these three, $\lambda=1$ is the only solution consistent
 with $h_0+\frac{1}{16}=\frac{3}{16}(1-2bQ)$.
 Going back to the analysis of the equation (\ref{nsFi}),
 we are thus lead to conjecture that after setting
 $\mu_{01}=\mu_{02}=0$ in (\ref{nsFe}),
 the function $F_0(\eta)$ should be given by
\begin{eqnarray}
  F_0(p_1,p_2,p_3;\eta) &=&
  \eta^{\frac{b\alpha_1}{2}+\frac{3}{8}}
  (1-\eta)^{\frac{b\alpha_2}{2}+\frac{3}{8}}
  \int dwdw'
  \left[w'(w'-1)(w'-\eta)\right]^{-\frac{3}{4}}
  \nonumber \\ && ~~\times
  w^{-\frac{3}{4}+\frac{ib}{2}p_{-1+2+3}}
  (w-1)^{-\frac{3}{4}+\frac{ib}{2}p_{1-2+3}}
  (w-\eta)^{-\frac{3}{4}-\frac{ib}{2}p_{1+2+3}}
  (w-w') \nonumber\\
  &\equiv&
  \eta^{\frac{b\alpha_1}{2}+\frac{3}{8}}
  (1-\eta)^{\frac{b\alpha_2}{2}+\frac{3}{8}}
  \int dwdw'f(w,w';\eta)(w-w'),
\end{eqnarray}
 with certain integration contours for $w$ and $w'$.
 Other three functions should be obtained from the symmetry
 of the equation (\ref{nsFi}): by flipping the signs of $p_{1,2,3}$
 and make a suitable change of coordinates we find a set of
 contour integrals
\begin{eqnarray}
  \cF_0(\eta) &=&
  \eta^{\frac{b\alpha_1}{2}+\frac{3}{8}}
  (1-\eta)^{\frac{b\alpha_2}{2}+\frac{3}{8}}
   \int dwdw'f(w,w';\eta)(w-w'),\nonumber\\
  \cF_1(\eta) &=&
  \eta^{\frac{b\alpha_1}{2}+\frac{3}{8}}
  (1-\eta)^{\frac{b\alpha_2}{2}-\frac{1}{8}}
   \int dwdw'f(w,w';\eta)(ww'-w-w'+\eta),\nonumber\\
  \cF_2(\eta) &=&
  \eta^{\frac{b\alpha_1}{2}-\frac{1}{8}}
  (1-\eta)^{\frac{b\alpha_2}{2}+\frac{3}{8}}
   \int dwdw'f(w,w';\eta)(ww'-\eta),\nonumber\\
  \cF_3(\eta) &=&
  \eta^{\frac{b\alpha_1}{2}-\frac{1}{8}}
  (1-\eta)^{\frac{b\alpha_2}{2}-\frac{1}{8}}
   \int dwdw'f(w,w';\eta)(ww'-w\eta-w'\eta+\eta),
\end{eqnarray}
 satisfying the following differential equations
\begin{eqnarray}
  2ib^{-1}{\cal D}\cF_0(\eta) &=&
 ~~p_1(1-\eta)^{1/2} \cF_1(\eta)
  +p_2 \eta^{1/2} \cF_2(\eta)
  -p_3 \eta^{1/2}(1-\eta)^{1/2}\cF_3(\eta),
     \nonumber \\
  2ib^{-1}{\cal D}\cF_1(\eta) &=&
 ~~p_1(1-\eta)^{1/2} \cF_0(\eta)
  +p_2 \eta^{1/2} \cF_3(\eta)
  -p_3 \eta^{1/2}(1-\eta)^{1/2}\cF_2(\eta),
     \nonumber \\
  2ib^{-1}{\cal D}\cF_2(\eta) &=&
  -p_1(1-\eta)^{1/2} \cF_3(\eta)
  +p_2 \eta^{1/2} \cF_0(\eta)
  +p_3 \eta^{1/2}(1-\eta)^{1/2}\cF_1(\eta),
     \nonumber \\
  2ib^{-1}{\cal D}\cF_3(\eta) &=&
  -p_1(1-\eta)^{1/2} \cF_2(\eta)
  +p_2 \eta^{1/2} \cF_1(\eta)
  +p_3 \eta^{1/2}(1-\eta)^{1/2}\cF_0(\eta),
     \nonumber \\
  {\cal D} &\equiv&
  \eta(1-\eta)\left[
    \partial_\eta + \frac{1-2bQ}{8\eta}+ \frac{1-2bQ}{8(\eta-1)}\right].
\end{eqnarray}
 Hence they can be used to express the solutions of (\ref{nsFi}).
 They are not yet functions because the contours are not specified.
 However, a notable property is that, as far as the form of
 the integrand is concerned, the four transform into one another
 under the change of integration variables.
 Some typical ones are given below:
\begin{eqnarray}
  w  \rightarrow 1-w,~~w' \rightarrow 1-w'  &:&
  ( \cF_0, \cF_1, \cF_2, \cF_3)(\eta) \rightarrow
  (-\cF_0, \cF_2, \cF_1, \cF_3)(1-\eta)_{(p_1,p_2)\rightarrow (p_2,p_1)},
 \nonumber \\
  w  \rightarrow \eta w ^{-1} &:&
  ( \cF_0, \cF_1, \cF_2, \cF_3)(\eta) \rightarrow
  ( \cF_2, \cF_3, \cF_0, \cF_1)(\eta)_{(p_1,p_3)\rightarrow -(p_1,p_3)},
 \nonumber \\
  w' \rightarrow \eta w'^{-1} &:&
  ( \cF_0, \cF_1, \cF_2, \cF_3)(\eta) \rightarrow
  ( \cF_2,-\cF_3, \cF_0,-\cF_1)(\eta).
\end{eqnarray}

   Now that we have found a way to express the solutions of the
 differential equation in an integral form, we have to investigate
 the property of them under the monodromy and find some formulae
 for the change of basis like those of hypergeometric functions.
 This can be done straightforwardly because the functions $\cF_i(\eta)$
 in the above are all expressible as sums of products of
 two hypergeometric functions.

   By fixing the integration contours and making
 a suitable rescaling, we define the function $\cF_0$
 in the following way:
\begin{eqnarray}
 \cF_0(p_1,p_2,p_3;\eta)
 &\equiv&
  \frac{8\Gamma(\frac{1}{2}+ibp_1)\Gamma(\frac{3}{2})}
       {\Gamma(\frac{1}{4}+\frac{ib}{2}p_{1+2-3})
        \Gamma(\frac{1}{4}+\frac{ib}{2}p_{1-2+3})
        \Gamma(\frac{1}{4})^2}
\nonumber \\ && \times
  \eta    ^{\frac{b\alpha_1}{2}+\frac{3}{8}}
  (1-\eta)^{\frac{b\alpha_2}{2}+\frac{3}{8}}
  \int_0^1dwdw'
  [w'(1-w')(1-w'\eta)]^{-\frac{3}{4}}
\nonumber \\ && \times
  w^{-\frac{3}{4}+\frac{ib}{2}p_{1+2-3}}
  (1-w)^{-\frac{3}{4}+\frac{ib}{2}p_{1-2+3}}
  (1-w\eta)^{-\frac{3}{4}-\frac{ib}{2}p_{1+2+3}}
  (w'-w).
\end{eqnarray}
   Using the same normalization and contour to define all the functions
 $\cF_i$, we find $\cF_0=\cF_1$ and $\cF_2=\cF_3$.
 Thus we introduce the following functions
\begin{eqnarray}
\lefteqn{
 -4\cG_0(p_1,p_2,p_3;\eta) ~\equiv~
   \cF_0(p_1,p_2,p_3;\eta) ~=~
   \cF_1(p_1,p_2,p_3;\eta) }\nonumber \\
 &=&
  \left(\frac{1+2ibp_{1-2+3}}{1+2ibp_1}\right)
  \eta    ^{\frac{b\alpha_1}{2}+\frac{3}{8}}
  (1-\eta)^{\frac{b\alpha_2}{2}+\frac{3}{8}}
  F(\tfrac{3}{4},\,\tfrac{5}{4},\,\tfrac{3}{2};\,\eta)
\nonumber \\ && ~\times
  F(\tfrac{3}{4}+\tfrac{ib}{2}p_{1+2+3},\,
    \tfrac{1}{4}+\tfrac{ib}{2}p_{1+2-3},\,
    \tfrac{3}{2}+ibp_1;\,\eta)
  -(p_{2,3}\rightarrow -p_{2,3}),
\nonumber \\
\lefteqn{
  2\cG_1(p_1,p_2,p_3;\eta) ~\equiv~
   \cF_2(p_1,p_2,p_3;\eta) ~=~
   \cF_3(p_1,p_2,p_3;\eta) }\nonumber \\
 &=&
  \left(\frac{1+2ibp_{1-2+3}}{1+2ibp_1}\right)
  \eta    ^{\frac{b\alpha_1}{2}-\frac{1}{8}}
  (1-\eta)^{\frac{b\alpha_2}{2}-\frac{1}{8}}
  F(-\tfrac{1}{4},\,\tfrac{1}{4},\,\tfrac{1}{2};\,\eta)
\nonumber \\ && ~\times
  F(\tfrac{3}{4}+\tfrac{ib}{2}p_{1+2+3},\,
    \tfrac{1}{4}+\tfrac{ib}{2}p_{1+2-3},\,
    \tfrac{3}{2}+ibp_1;\,\eta)
  +(p_{2,3}\rightarrow -p_{2,3})
\end{eqnarray}
 We furthermore introduce the notations
\begin{equation}
  \cG_2( p_1, p_2, p_3;\eta) = 
  \cG_0(-p_1, p_2,-p_3;\eta),~~~
  \cG_3( p_1, p_2, p_3;\eta) = 
  \cG_1(-p_1, p_2,-p_3;\eta),
\end{equation}
 so that the solution of the differential equations for
 $f^{\ep_1\ep_2\ep_3\ep_4}$ of (\ref{SSSS}) be
 linear combinations of the functions $\cG_i( p_1, p_2, p_3;\eta)$
 or of $\cG_i( p_1, p_2,-p_3;\eta)$.
 Note that the hypergeometric functions in the above which do
 not depend on $p_i$ can also be written as
\begin{eqnarray}
  F(\tfrac{3}{4},\tfrac{5}{4},\tfrac{3}{2};\eta)
 &=& (1-\eta)^{-1/2}\left(\frac{2}{1+\sqrt{1-\eta}}\right)^{1/2}, \nonumber\\
  F(-\tfrac{1}{4},\tfrac{1}{4},\tfrac{1}{2};\eta)
 &=& \left(\frac{1+\sqrt{1-\eta}}{2}\right)^{1/2},
\end{eqnarray}
 so that they are related to the four-point functions of
 spin operators in Ising model\cite{BPZ}.
 The above functions ${\cal G_i}$ are not single valued
 on the entire ${\bf CP}^1$,
 and the basis of solutions are chosen so as to diagonalize the
 monodromy around $\eta=0$.
 Our next task is to find the transition coefficients giving
 a relation between different bases.
 Using the formula for hypergeometric functions
 they are given by
\begin{equation}
\left(\begin{array}{c}
  \cG_0(p_1,p_2,p_3;\eta) \\
  \cG_1(p_1,p_2,p_3;\eta) \\
  \cG_2(p_1,p_2,p_3;\eta) \\
  \cG_3(p_1,p_2,p_3;\eta)
\end{array}\right)
 = \frac{1}{\sqrt{2}}
   \left(\begin{array}{rrrr}
    -x_{+-} &-x_{+-} & x_{++} & x_{++} \\
    -x_{+-} & x_{+-} &-x_{++} & x_{++} \\
     x_{--} &-x_{--} &-x_{-+} & x_{-+} \\
     x_{--} & x_{--} & x_{-+} & x_{-+}
   \end{array}\right)
\left(\begin{array}{c}
  \cG_0(p_2,p_1,p_3;1-\eta) \\
  \cG_1(p_2,p_1,p_3;1-\eta) \\
  \cG_2(p_2,p_1,p_3;1-\eta) \\
  \cG_3(p_2,p_1,p_3;1-\eta)
\end{array}\right)
\end{equation}
\begin{eqnarray}
  x_{++} &=&
 \frac{\Gamma(\frac{1}{2}+ibp_1)\Gamma(\frac{1}{2}+ibp_2)}
      {\Gamma(\frac{3}{4}+\frac{ib}{2}p_{1+2+3})
       \Gamma(\frac{1}{4}+\frac{ib}{2}p_{1+2-3})},\nonumber \\
  x_{+-} &=&
 \frac{\Gamma(\frac{1}{2}+ibp_1)\Gamma(\frac{1}{2}-ibp_2)}
      {\Gamma(\frac{3}{4}+\frac{ib}{2}p_{1-2-3})
       \Gamma(\frac{1}{4}+\frac{ib}{2}p_{1-2+3})},\nonumber \\
  x_{-+} &=&
 \frac{\Gamma(\frac{1}{2}-ibp_1)\Gamma(\frac{1}{2}+ibp_2)}
      {\Gamma(\frac{3}{4}+\frac{ib}{2}p_{-1+2-3})
       \Gamma(\frac{1}{4}+\frac{ib}{2}p_{-1+2+3})},\nonumber \\
  x_{--} &=&
 \frac{\Gamma(\frac{1}{2}-ibp_1)\Gamma(\frac{1}{2}-ibp_2)}
      {\Gamma(\frac{3}{4}+\frac{ib}{2}p_{-1-2+3})
       \Gamma(\frac{1}{4}+\frac{ib}{2}p_{-1-2-3})}.
\end{eqnarray}
   Using the above formulae we shall then try to find monodromy
 invariant combinations of the left and right sectors.
 First, there is a ``diagonal'' product of the following form:
\begin{eqnarray}
 &&
  \left[
  x_{--}x_{-+}\left(\cG_0\bar{\cG}_0+\cG_1\bar{\cG}_1\right)
 +x_{+-}x_{++}\left(\cG_2\bar{\cG}_2+\cG_3\bar{\cG}_3\right)
  \right](p_1,p_2,p_3;\eta)
\nonumber \\&& ~=~
  \left[
  x_{--}x_{+-}\left(\cG_0\bar{\cG}_0+\cG_1\bar{\cG}_1\right)
 +x_{-+}x_{++}\left(\cG_2\bar{\cG}_2+\cG_3\bar{\cG}_3\right)
  \right](p_2,p_1,p_3;1-\eta)
\label{S4d}
\end{eqnarray}
 where we denoted $\bar{\cG}_i(p_i;\eta)\equiv\cG_i(p_i;\bar{\eta})$.
 Up to an overall constant, it has the following asymptotic behavior
 at $\eta\sim0$:
\begin{eqnarray}
   \left(\frac{ibp_{2-3}}{1+2ibp_1}\right)^2
   |\eta|^{b\alpha_1+\frac{3}{4}}
  +|\eta|^{b(Q-\alpha_1)-\frac{1}{4}}
   \gamma(\tfrac{1}{2}+ibp_1)^2
   \gamma(\tfrac{1}{4}+\tfrac{ib}{2}p_{-1+2+3})
   \gamma(\tfrac{1}{4}+\tfrac{ib}{2}p_{-1-2-3})
 \nonumber \\ \times
   \gamma(\tfrac{3}{4}+\tfrac{ib}{2}p_{-1+2-3})
   \gamma(\tfrac{3}{4}+\tfrac{ib}{2}p_{-1-2+3})
 \nonumber \\
  +|\eta|^{b\alpha_1-\frac{1}{4}}
  +|\eta|^{b(Q-\alpha_1)+\frac{3}{4}}
   \left(\frac{ibp_{2+3}}{1-2ibp_1}\right)^2
   \gamma(\tfrac{1}{2}+ibp_1)^2
   \gamma(\tfrac{1}{4}+\tfrac{ib}{2}p_{-1+2+3})
   \gamma(\tfrac{1}{4}+\tfrac{ib}{2}p_{-1-2-3})
 \nonumber \\ \times
   \gamma(\tfrac{3}{4}+\tfrac{ib}{2}p_{-1+2-3})
   \gamma(\tfrac{3}{4}+\tfrac{ib}{2}p_{-1-2+3})
\label{asS4}
\end{eqnarray}

   Let us compare the asymptotic behavior of the above solution
 with the previous OPE analysis based on the assumption
 that $\Theta_{-b/2}^{\ep\bep}\Theta_\alpha^{\tau\btau}$
 is expanded into two discrete terms.
 Apart from the coordinate dependences which are irrelevant,
 the four-point functions should obey the following asymptotic behavior:
\begin{eqnarray}
\lefteqn{
  \vev{\Theta_{-b/2}^{--}(z_0)
       \Theta_{\alpha_1}^{++}(z_1)
       \Theta_{\alpha_2}^{++}(z_2)
       \Theta_{\alpha_3}^{++}(z_3)} }
 \nonumber \\
 &\sim&
  |z_{01}|^{b\alpha_1+\frac{3}{4}}
  \left(\frac{ibp_{2-3}}{1+2ibp_1}\right)^2
  \tilde{C}_1(\alpha_1-\tfrac{b}{2};\alpha_2,\alpha_3)
 +|z_{01}|^{b(Q-\alpha_1)-\frac{1}{4}}
   \frac{\tilde{C}_-(\alpha_1)
         \tilde{C}_3(\alpha_1+\tfrac{b}{2};\alpha_2,\alpha_3)}{2i}
 \nonumber \\
\lefteqn{
  \vev{\Theta_{-b/2}^{++}(z_0)
       \Theta_{\alpha_1}^{++}(z_1)
       \Theta_{\alpha_2}^{++}(z_2)
       \Theta_{\alpha_3}^{++}(z_3)} }
 \nonumber \\
 &\sim&
  |z_{01}|^{b\alpha_1-\frac{1}{4}}
        \tilde{C}_3(\alpha_1-\tfrac{b}{2};\alpha_2,\alpha_3)
 +|z_{01}|^{b(Q-\alpha_1)+\frac{3}{4}}
  \left(\frac{ibp_{2-3}}{1-2ibp_1}\right)^2
  \frac{\tilde{C}_-(\alpha_1)
        \tilde{C}_1(\alpha_1+\tfrac{b}{2};\alpha_2,\alpha_3)}
       {2i}
\end{eqnarray}
 Comparing them with (\ref{asS4}) we find that
 the crossing symmetric solution (\ref{S4d}) of the differential equation
 corresponds to the ``sum'' of four-point functions
\begin{eqnarray}
\lefteqn{
  x_{--}x_{-+}(\cG_0\bar{\cG}_0+\cG_1\bar{\cG}_1)
 +x_{+-}x_{++}(\cG_2\bar{\cG}_2+\cG_3\bar{\cG}_3)}
 \nonumber \\  &\sim&
    \vev{\Theta_{-b/2}^{--}
         \Theta_{\alpha_1}^{++}
         \Theta_{\alpha_2}^{++}
         \Theta_{\alpha_3}^{++}}
   +\vev{\Theta_{-b/2}^{++}
         \Theta_{\alpha_1}^{--}
         \Theta_{\alpha_2}^{++}
         \Theta_{\alpha_3}^{++}}
 \nonumber \\ && \hskip-4mm
   +\vev{\Theta_{-b/2}^{++}
         \Theta_{\alpha_1}^{++}
         \Theta_{\alpha_2}^{--}
         \Theta_{\alpha_3}^{++}}
   +\vev{\Theta_{-b/2}^{++}
         \Theta_{\alpha_1}^{++}
         \Theta_{\alpha_2}^{++}
         \Theta_{\alpha_3}^{--}}.
\end{eqnarray}

   By flipping the momentum of one of the three generic spin fields
 we obtain another diagonal type solution of the differential equation,
 and these two are the only diagonal solutions we could find.
 One might think that a more careful analysis would lead to
 another solution and thereby enable us to write down
 the expression for each of the summand in the above.
 However, there should not be any more solutions once we admit that
 all the solutions are expressed in terms of
 $\cG_i(p_i;\eta)$.
 It is also difficult to argue that the solutions be duplicated
 due to the ambiguity in choosing the square-root branches.
 Thus we conclude that there are only two independent solutions
 of diagonal type.

   The above result seems to indicate that in correlation functions
 involving spin fields, all one can fix by hand is the total chirality
 of the product of spin fields and not the chirality of each spin field.
 Nevertheless, our previous analysis of two- and three-point
 structure constants still remains valid since they involve
 no more than two spin fields.

   Finally, there are also solutions of the off-diagonal type:
\begin{eqnarray}
\lefteqn{
 \left[
  x_{--}x_{-+}(\cG_0\bar{\cG}_1-\cG_1\bar{\cG}_0)
 -x_{+-}x_{++}(\cG_2\bar{\cG}_3-\cG_3\bar{\cG}_2)\right](p_1,p_2,p_3;\eta)
}\nonumber \\
 &=&
 -\left[
  x_{--}x_{+-}(\cG_0\bar{\cG}_1-\cG_1\bar{\cG}_0)
 -x_{-+}x_{++}(\cG_2\bar{\cG}_3-\cG_3\bar{\cG}_2)\right](p_2,p_1,p_3;1-\eta)
 \nonumber \\  &\sim&
    \vev{\Theta_{-b/2}^{--}
         \Theta_{\alpha_1}^{+-}
         \Theta_{\alpha_2}^{+-}
         \Theta_{\alpha_3}^{++}}
   +\vev{\Theta_{-b/2}^{++}
         \Theta_{\alpha_1}^{-+}
         \Theta_{\alpha_2}^{+-}
         \Theta_{\alpha_3}^{++}}
 \nonumber \\ && \hskip-4mm
   +\vev{\Theta_{-b/2}^{++}
         \Theta_{\alpha_1}^{+-}
         \Theta_{\alpha_2}^{-+}
         \Theta_{\alpha_3}^{++}}
   +\vev{\Theta_{-b/2}^{++}
         \Theta_{\alpha_1}^{+-}
         \Theta_{\alpha_2}^{+-}
         \Theta_{\alpha_3}^{--}}.
\end{eqnarray}

\section{Super Liouville Theory with Boundary}
   If the worldsheet has a boundary, the action has boundary terms:
\begin{eqnarray}
  I&=&\frac{1}{2\pi}\int_\Sigma d^2z\left[
       \partial\phi\bar{\partial}\phi +\tfrac{QR\phi}{4}
      +\psi\bar{\partial}\psi+\bar{\psi}\partial\bar{\psi}
      \right]
      +2i\mu b^2\int_\Sigma d^2z \psi\bar{\psi}e^{b\phi}
  \nonumber \\ && ~~
   + \oint_{\partial\Sigma}dx\left[
       \tfrac{QK\phi}{4\pi}+a\mu_B b\psi e^{b\phi/2}
     \right],
\end{eqnarray}
 where $K$ is the curvature of the boundary
 which is defined by the Euler number formula
\begin{equation}
  \frac{1}{4\pi}\int_{\Sigma}\sqrt{g}R
 +\frac{1}{2\pi}\int_{\partial\Sigma}g^{1/4}K
 =\chi=2-2g-h
\end{equation}
 for worldsheets $\Sigma$ with $g$ handles and $h$ holes.
 The coupling $\mu_B$ will be referred to as the boundary cosmological
 constant as in the bosonic case.
 It can take different values for different connected components
 of the boundary, and it may also jump at points where
 the boundary primary operators are inserted.
 The boundary interaction term contains a Grassmann odd constant
 $a$ satisfying $a^2=1$, in order to avoid the Lagrangian becoming
 Grassmann odd\cite{GZ,N,ARS}.

   On worldsheets with boundary we can insert operators on the boundary.
 One of the most fundamental boundary operator is
\begin{equation}
  B_\beta(x) = e^{\beta\phi/2}(x) = e^{\beta\phi_L}(x),
\end{equation}
 and of course there are some other operators like
\begin{equation}
  \psi B_\beta = \psi e^{\beta\phi_L},~~~
  \Theta_\beta^\pm = \sigma^\pm e^{\beta\phi_L}.
\end{equation}
 In free-field scheme the correlators of bulk- and boundary fields
 on the upper half-plane can be calculated as usual using
 mirror image techniques.
 One can also perform the integration over the zero-mode of $\phi$
 and derive that any correlator diverges when
\begin{equation}
  Q(2-2g-h)-2\sum_{\makebox[0mm][l]{\tiny ~~~ $i$ (bulk)}}\alpha_i
          ~-~\sum_{\makebox[0mm][l]{\tiny ~~~ $j$ (boundary)}}\beta_j
  ~=~ b(2n+n_B)
\end{equation}
 for non-negative integers $n$ and $n_B$, and the residue is given by
 a sum of free field correlators with $n$ bulk and $n_B$ boundary
 screening operators.

\subsection{Classification of boundary states}

   The classification of boundary states can be done by studying
 the modular property of annulus partition functions.
 We follow the same path as the reference \cite{ZZ} which
 analyzed the bosonic Liouville theory.
 We first introduce the Ishibashi states, and then classify all
 the possible Cardy states by expressing them as superpositions
 of Ishibashi states.
 Note that in super Liouville theory there is a freedom in choosing
 the spin structure on the worldsheet, so that we have to consider
 the characters with $(-)^{F}$ inserted as well as the ordinary ones.
 
   Boundary states in generic superconformal field theory are
 defined as the solutions of the boundary condition on currents
 on the real axis:
\begin{equation}
  T(z)=\bar{T}(\bar{z}),~~~
  T_F(z)=\zeta\bar{T}_F(\bar{z}).
\end{equation}
 Mapping the upper half-plane onto a unit disc the above condition
 can be rewritten into the following form
\begin{equation}
  L_n=\bar{L}_{-n},~~~
  G_r=-i\zeta\bar{G}_{-r}.
\end{equation}
 The boundary states $\ket{B;\zeta}$ and $\bra{B;\zeta}$ are
 therefore solutions of the equations
\begin{equation}
\begin{array}{rcrcl}
  \bra{B;\zeta}(L_n-\bar{L}_{-n})&=&
  \bra{B;\zeta}(G_r+i\zeta\bar{G}_{-r})&=&0,\\
  (L_n-\bar{L}_{-n})\ket{B;\zeta}&=&
  (G_r-i\zeta\bar{G}_{-r})\ket{B;\zeta}&=&0.
\end{array}
\end{equation}
 The NS (R) boundary states satisfy the above condition with
 $r\in{\bf Z}+\frac{1}{2}$ ($r\in{\bf Z}$).

 Given a highest weight state $\ket{h;\NS}$ of
 the superconformal algebra, one can construct the Ishibashi state
 $\iket{h;+,\zeta}$ in the following way:
\begin{equation}
  \iket{h;+,\zeta}=\ket{h;\NS}_L\ket{h;\NS}_R + (\mbox{descendants})
\end{equation}
 In the same way one can construct the Ishibashi state
 $\iket{h;-,\zeta}$ from a highest weight state $\ket{h;\R^\pm}$
 in the R sector:
\begin{eqnarray}
  \iket{h;-,\zeta}&=&
  \textstyle
  \ket{h;\R^+}_L\ket{h;\R^+}_R -i\zeta
  \ket{h;\R^-}_L\ket{h;\R^-}_R +
  (\mbox{descendants}), \nonumber \\
  \ibra{h;-,\zeta}&=&
  \textstyle
  \bra{h;\R^+}_R\bra{h;\R^+}_L +i\zeta
  \bra{h;\R^-}_R\bra{h;\R^-}_L +
  (\mbox{descendants}),
\label{IR}
\end{eqnarray}
 where we assumed
 $\vev{h;\R^+|h;\R^+}=\vev{h;\R^-|h;\R^-}$ to be nonzero.
 Note that we will only consider the combinations of even total
 chirality in what follows.
 From these definitions it follows that the annulus partition function
 bounded by two Ishibashi states is given by the character:
\begin{eqnarray}
  \ibra{h;+,\zeta}e^{i\pi\tau_c(L_0+\bar{L}_0-\frac{c}{12})}
  \iket{k;+,\zeta'}
 &=& \delta_{h,k}{\rm Tr}_{h(\NS)}
   [e^{2i\pi\tau_c(L_0-\frac{c}{24})}
    (\zeta\zeta')^F], \nonumber\\
  \ibra{h;-,\zeta}e^{i\pi\tau_c(L_0+\bar{L}_0-\frac{c}{12})}
  \iket{k;-,\zeta'}
 &=& \delta_{h,k}{\rm Tr}_{h(\R)}
   [e^{2i\pi\tau_c(L_0-\frac{c}{24})}
   (\zeta\zeta')^F].
\end{eqnarray}
 Here the expression is symbolic in the sense that
 the delta symbol $\delta_{h,k}$  merely represents
 the Ishibashi states diagonalize the annulus partition function
 as seen from the closed string channel.
 R Ishibashi states and NS Ishibashi states are orthogonal to each
 other.

   Cardy states are defined by the property that
 the multiplicity of open string modes between two of them is
 given by the fusion coefficient ${\cal N}_{h,k}^{~l}$.
 In superconformal field theories they are given by the sum
 of NS and R pieces:
\begin{equation}
  \cbra{h;\zeta} = \cbra{h;+,\zeta}+ \cbra{h;-,\zeta},~~~~
  \cket{h;\zeta} = \cket{h;+,\zeta}+ \cket{h;-,\zeta}.
\end{equation}
 NS (R) piece of any Cardy state is itself a solution of boundary
 condition on currents, and should be expressed as a superposition
 of NS (R) Ishibashi states.
 The partition function on an annulus bounded by two of them
 as seen from the open string channel is expressed as the sum
 of characters with coefficients ${\cal N}_{h,k}^{~l}$
\begin{equation}
  Z_{(h,\pm,\zeta),(k,\pm,\zeta')}(\tau_o)~=~
\left\{
\begin{array}{ll}
 {\cal N}_{h,k}^{~l}{\rm Tr}_{l(\NS)}
   [e^{2i\pi\tau_o(L_0-\frac{c}{24})}(\pm 1)^F] & (\zeta=\zeta'), \\
 {\cal N}_{h,k}^{~l}{\rm Tr}_{l(\R)}
   [e^{2i\pi\tau_o(L_0-\frac{c}{24})}(\pm 1)^F] & (\zeta=-\zeta').
\end{array}
\right.
\end{equation}

   Let us then consider the boundary states in super Liouville theory.
 First of all, as the labels $h$ of representations of
 the superconformal algebra, we use the Liouville momentum $p$
 for non-degenerate representations or the index $(r,s)$ for degenerate ones.
 The characters for non-degenerate representations are
 given by ($q\equiv e^{2\pi i\tau}$)
\begin{equation}
\begin{array}{rclcl}
  {\rm Tr}_{p(\NS)}q^{L_0-\frac{c}{24}}
  &=& \chi_{p(\NS)}^+(\tau)
  &=& q^{\frac{p^2}{2}-\frac{1}{16}}
      \prod_{n=1}(1-q^{n})^{-1}(1+q^{n-\frac{1}{2}}), \\
  {\rm Tr}_{p(\NS)}(-)^Fq^{L_0-\frac{c}{24}}
  &=& \chi_{p(\NS)}^-(\tau)
  &=& q^{\frac{p^2}{2}-\frac{1}{16}}
      \prod_{n=1}(1-q^{n})^{-1}(1-q^{n-\frac{1}{2}}), \\
  {\rm Tr}_{p(\R)}q^{L_0-\frac{c}{24}}
  &=& \chi_{p(\R)}^+(\tau)
  &=& 2q^{\frac{p^2}{2}}
      \prod_{n=1}(1-q^{n})^{-1}(1+q^n).
\end{array}
\end{equation}
 Note that ${\rm Tr}_\R(-)^Fq^{L_0-\frac{c}{24}}$ vanishes
 for any representations except for the R vacuum. 
 They obey the following modular transformation property
\begin{equation}
\begin{array}{rcl}
  \chi_{u(\NS)}^+(\tau)
  &=& \mint_{-\infty}^\infty dp e^{2i\pi pu}\chi_{p(\NS)}^+(-1/\tau), \\
  2^{1/2}\chi_{u(\NS)}^-(\tau)
  &=& \mint_{-\infty}^\infty dp e^{2i\pi pu}\chi_{p(\R)}^+(-1/\tau), \\
  2^{-1/2}\chi_{u(\R)}^+(\tau)
  &=& \mint_{-\infty}^\infty dp e^{2i\pi pu}\chi_{p(\NS)}^-(-1/\tau).
\end{array}
\end{equation}
 For degenerate representations, the characters are given by
 those of corresponding Verma modules subtracted by those
 of null submodules:
\begin{equation}
\begin{array}{rclcl}
    \chi_{r,s(\NS)}^+
&=& \chi_{\frac{i}{2}(rb+sb^{-1})~(\NS)}^+
   -\chi_{\frac{i}{2}(rb-sb^{-1})~(\NS)}^+, \\
    \chi_{r,s(\NS)}^-
&=& \chi_{\frac{i}{2}(rb+sb^{-1})~(\NS)}^-
   -(-)^{rs}\chi_{\frac{i}{2}(rb-sb^{-1})~(\NS)}^-, \\
    \chi_{r,s(\R)}^+
&=& \chi_{\frac{i}{2}(rb+sb^{-1})~(\R)}^-
   -\chi_{\frac{i}{2}(rb-sb^{-1})~(\R)}^-.
\end{array}
\end{equation}

   We would like to find the expression for the wave functions
 $\Psi(p;h_\zeta)$ which express the NS and R Cardy states as superpositions
 of Ishibashi states belonging to normalizable representations:
\begin{eqnarray}
  \cbra{h;\pm,\zeta}
  &=& 2\int_0^\infty\frac{dp}{2\pi}\Psi_\pm(p;h_\zeta)  \ibra{p;\pm,\zeta},
 \nonumber \\
  \cket{h;\pm,\zeta}
  &=& 2\int_0^\infty\frac{dp}{2\pi}\iket{p;\pm,\zeta}\Psi_\pm^\dag(p;h_\zeta).
\end{eqnarray}
 Here we have taken care for the equivalence of representations
 with momentum $p$ and $-p$.
 The Ishibashi states are normalized to satisfy
\begin{equation}
\begin{array}{rcl}
 \ibra{p,+,\zeta}e^{i\pi\tau_c(L_0+\bar{L}_0-\frac{c}{12})}
 \iket{p',+,\zeta'}
 &=& 2\pi\delta(p-p')\chi_{p(\NS)}^{\zeta\zeta'}(\tau_c), \\
 \ibra{p,-,\zeta}e^{i\pi\tau_c(L_0+\bar{L}_0-\frac{c}{12})}
 \iket{p',-,\zeta'}
 &=& \sqrt{2}\pi\delta(p-p')\chi_{p(\R)}^{\zeta\zeta'}(\tau_c).
\end{array}
\end{equation}
 We also assume that $\Psi_\pm^\dag(p;h_\zeta)=\Psi_\pm(-p;h_\zeta)$.

   There is one important notice regarding the equivalence of
 R Ishibashi states under the reflection $p\rightarrow -p$.
 Recall that the highest weight states in the R sector can be
 created by multiplying a spin operator onto the vacuum.
 Therefore, if we write the R Ishibashi states as in (\ref{IR}),
 we obtain
\begin{eqnarray}
  \ibra{p;-,\zeta} &=&
  \bra{\Theta_{Q/2+ip}^{++}}
 -\zeta\bra{\Theta_{Q/2+ip}^{--}}
 +(\mbox{descendants}), \nonumber \\
  \iket{p;-,\zeta} &=&
  \ket{\Theta_{Q/2-ip}^{++}}
 -\zeta\ket{\Theta_{Q/2-ip}^{--}}
 +(\mbox{descendants}).
\end{eqnarray}
 As a consequence, if $\zeta=+1$ there arises a minus sign in
 flipping the sign of $p$.
 Hence the wave functions for R Cardy states should depend also
 on how the supercurrents in the left- and the right sectors are
 glued: they must be odd functions of $p$ for $\zeta=1$ and
 even functions for $\zeta=-1$.
 On the other hand, there is no such subtleties for NS Ishibashi states
 so that one may well expect that the NS wave functions do not
 depend on $\zeta$.

   The open/closed duality for annulus partition functions 
 together with the obvious fusion relation
 ${\cal N}_{(1,1),h}^{~~k}=\delta_h^k$ yields
\begin{eqnarray}
  \Psi_+(p;u_\zeta)\Psi_+(-p;(1,1)_\zeta) &=&
  \pi\cos(2\pi pu), \nonumber \\
  \Psi_+(p;(r,s)_\zeta)\Psi_+(-p;(1,1)_\zeta) &=&
  2\pi\sinh(\pi prb)\sinh(\pi ps/b),\nonumber \\
  \Psi_-(p;u_\zeta)\Psi_-(-p;(1,1)_\zeta) &=&
  \pi\cos(2\pi pu), \nonumber \\
  \Psi_-(p;(r,s)_\zeta)\Psi_-(-p;(1,1)_\zeta) &=&
  2\pi\sinh(\pi prb+\tfrac{i\pi rs}{2})
   \sinh(\tfrac{\pi ps}{b}-\tfrac{i\pi rs}{2})
\end{eqnarray}
 Another equation comes from the fact that the wave functions
 $\Psi_\pm(p;h_\zeta)$ are proportional to the disc one-point
 functions of $V_{\frac{Q}{2}+ip}$ or $\Theta_{\frac{Q}{2}+ip}$.
 Therefore they must be consistent with the reflection relations:
\begin{equation}
 \Psi_+(p;h_\zeta)=D(\tfrac{Q}{2}+ip)\Psi_+(-p;h_\zeta),~~~~
 \Psi_-(p;h_\zeta)=-\zeta\tilde{D}(\tfrac{Q}{2}+ip)\Psi_-(-p;h_\zeta).
\end{equation}
 These equations determine the form of almost all the wave functions:
\begin{eqnarray}
  \Psi_+(p;(1,1)_\zeta) &=&
 2^{\frac{1}{2}}\pi^{\frac{3}{2}}(\mu\pi\gamma(\tfrac{bQ}{2}))^{-ip/b}
 \left\{ip\Gamma(-ipb)\Gamma(-\tfrac{ip}{b})\right\}^{-1}     , \nonumber\\
  \Psi_+(p;(r,s)_\zeta) &=&
 -2^{\frac{1}{2}}\pi^{-\frac{1}{2}}(\mu\pi\gamma(\tfrac{bQ}{2}))^{-ip/b}\cdot
  ip\Gamma(ipb)\Gamma(\tfrac{ip}{b})
  \sinh(\pi prb)\sinh(\pi ps/b),
  \nonumber\\
  \Psi_+(p;u_\zeta) &=&
 -2^{-\frac{1}{2}}\pi^{-\frac{1}{2}}(\mu\pi\gamma(\tfrac{bQ}{2}))^{-ip/b}\cdot
  ip\Gamma(ipb)\Gamma(\tfrac{ip}{b})\cos(2\pi pu), \nonumber\\
  \Psi_-(p;(1,1)_-) &=&
 2^{\frac{1}{2}}\pi^{\frac{3}{2}}(\mu\pi\gamma(\tfrac{bQ}{2}))^{-ip/b}
 \left\{\Gamma(\tfrac{1}{2}-ipb)\Gamma(\tfrac{1}{2}-\tfrac{ip}{b})
 \right\}^{-1}, \nonumber\\
  \Psi_-(p;(r,s)_-) &=&
 2^{\frac{1}{2}}\pi^{-\frac{1}{2}}(\mu\pi\gamma(\tfrac{bQ}{2}))^{-ip/b}
  \Gamma(\tfrac{1}{2}+ipb)
  \Gamma(\tfrac{1}{2}+\tfrac{ip}{b})
 \nonumber \\ && \hskip30mm\times
   \sinh(\pi prb+\tfrac{i\pi rs}{2})
   \sinh(\tfrac{\pi ps}{b}-\tfrac{i\pi rs}{2}),
  \nonumber\\
  \Psi_-(p;u_-) &=&
 2^{-\frac{1}{2}}\pi^{-\frac{1}{2}}(\mu\pi\gamma(\tfrac{bQ}{2}))^{-ip/b}
  \Gamma(\tfrac{1}{2}+ipb)\Gamma(\tfrac{1}{2}+\tfrac{ip}{b})\cos(2\pi pu).
\end{eqnarray}
 All we are left with is the R wave functions $\Psi_-(p;h_+)$.
 However, due to the requirement that they must be a product of
 Gamma functions multiplied by an {\it odd} function of $p$,
 one can actually find no analytic expressions for them.
 Consequently, one should conclude that the R 
 wave functions cannot be found by analyzing the modular property.
 They will be proposed later by an analysis of one-point function
 on a disc and, moreover, we will find that the degenerate Cardy
 states $(r,s)_\zeta$ must satisfy $r+s={\rm even(odd)}$ for
 $\zeta=-1(+1)$.
 If we accept this, the absence of $(1,1)_+$ state explains
 why we could not find the wave functions the Cardy states with
 $\zeta=1$ from the modular property.

   It might seem strange that one can find the R Cardy states
 with $\zeta=-1$ only, because naively one tends to think that the two
 choices for the boundary conditions on supercurrent should be
 equivalent.
 However, it turned out that the two choices are actually
 inequivalent and affects the parity of the wave functions
 under the sign-change of the momentum $p$.

\subsection{One-point functions of bulk operators}

   Let us try to reproduce these wave functions from a different
 approach, by calculating the one-point functions on a disc.
 We define various one-point structure constants by the equations
\begin{eqnarray}
  \vev{V_\alpha(z)}_{u_\zeta}
 &=& U_+(\alpha;u_\zeta)|z-\bar{z}|^{-2h_\alpha},
\nonumber \\
  \vev{\Theta_\alpha^{\ep\ep}(z)}_{u_\zeta}
 &=& U_-(\alpha,\ep;u_\zeta)|z-\bar{z}|^{-2h_\alpha-\frac{1}{8}}.
\end{eqnarray}
 The one-point functions of spin fields $\vev{\Theta_\alpha^{\ep,-\ep}}$
 always vanish because we restricted the R boundary states
 to have even total chirality in (\ref{IR}).
 The periodicity of supercurrents when we go around the boundary
 of the disc is unambiguously determined by how many spin fields are
 inserted on the disc.
 All the other one-point functions are zero or obtained 
 by superconformal transformations from the above ones.
 For example, the one-point function of descendants
 in the NS sector is given by
\begin{equation}
  \vev{\psi\bar{\psi}V_\alpha(z)}_{u_\zeta}
 = i\zeta\cdot(Q-\alpha)\alpha^{-1}
   U_+(\alpha;u_\zeta)|z-\bar{z}|^{-2h_\alpha-1},
\end{equation}
 and the one-point functions of spin fields depend on the index $\ep$
 in the following way:
\begin{equation}
  \vev{\Theta_\alpha^{\ep\ep}(z)}_{u_\zeta}
 =-\zeta\vev{\Theta_\alpha^{-\ep,-\ep}(z)}_{u_\zeta},
\label{SeTf}
\end{equation}
 in consistency with the previous analysis of modular property.

   To obtain the one-point structure constants, we derive
 a set of recursion relations for them from the solution
 of differential equation for two-point functions with
 one degenerate operator.
 Let us first consider
\begin{eqnarray}
\lefteqn{
  \vev{V_\alpha(z)\Theta_{-b/2}^{\ep\ep}(w)}_{u_\zeta}
~=~|w-\bar{z}|^{-4h_{-b/2}-\frac{1}{4}}
   |z-\bar{z}|^{2h_{-b/2}-2h_\alpha-\frac{1}{8}}\times }
 \nonumber \\ && \hskip-6mm
 \times\left\{
   C_+(\alpha)U_-(\alpha-\tfrac{b}{2},\ep;u_\zeta)
   \eta^{\frac{b\alpha}{2}}(1-\eta)^{-\frac{b^2}{4}+\frac{3}{8}}
   F(\tfrac{1-b^2}{2}+ibp_\alpha,\,\tfrac{1-b^2}{2},\,1+ibp_\alpha;\eta)
        \right. \nonumber \\ && \left. \hskip-11mm
  +C_-(\alpha)U_-(\alpha+\tfrac{b}{2},-\ep;u_\zeta)
   \eta^{\frac{b(Q-\alpha)}{2}}(1-\eta)^{-\frac{b^2}{4}+\frac{3}{8}}
   F(\tfrac{1-b^2}{2}-ibp_\alpha,\,\tfrac{1-b^2}{2},\,1-ibp_\alpha;\eta)
   \right\},
\end{eqnarray}
 where $\eta=\frac{|z-w|^2}{|z-\bar{w}|^2}$ and $C_\pm(\alpha)$
 are the OPE coefficients defined in (\ref{CNS}).
 The coefficients are chosen so that the behavior
 when the two bulk operators approach each other
 agrees with the OPE analysis.
 On the other hand, when $\Theta_{-b/2}^{\ep\ep}$ approaches the
 boundary, it should be expanded as a discrete sum of boundary
 degenerate operators:
\begin{equation}
  \Theta_{-b/2}^{\ep\ep}(w)\rightarrow
  |w-\bar{w}|^{-2h_{-b/2}+h_{-b}+\frac{3}{8}}r_+\psi B_{-b}(w)
 +|w-\bar{w}|^{-2h_{-b/2}-\frac{1}{8}}r_-B_0(w)
\end{equation}
 with certain coefficients $r_\pm$.
 Comparing this with the behavior of the solution around $\eta\sim1$
 we obtain a recursion relation:
\begin{eqnarray}
   C_+(\alpha)U_-(\alpha-\tfrac{b}{2},\ep;u_\zeta)
  \frac{\Gamma(b\alpha+\frac{1-b^2}{2})\Gamma(-b^2)}
       {\Gamma(b\alpha-b^2)\Gamma(\frac{1-b^2}{2})}
 ~~~~
 \nonumber \\
  +C_-(\alpha)U_-(\alpha+\tfrac{b}{2},-\ep;u_\zeta)
  \frac{\Gamma(\frac{3+b^2}{2}-b\alpha)\Gamma(-b^2)}
       {\Gamma(1-b\alpha)\Gamma(\frac{1-b^2}{2})}
 &=& r_-(\ep,\zeta)U_+(\alpha;u_\zeta).
\label{rcU1}
\end{eqnarray}
 The coefficient $r_-$ can be calculated using free fields
\begin{equation}
  r_- = -ab\mu_B|w-\bar{w}|^{-\frac{3b^2}{4}-\frac{3}{8}}
        \int dx\vev{\Theta_{-b/2}^{\ep\ep}(w)\psi B_b(x)B_Q(y)}
      =-\sqrt{2}\pi\hat{r} b\mu_B\Gamma(-b^2)\Gamma(\tfrac{1-b^2}{2})^{-2},
\end{equation}
 where $\hat{r}$ is related to the free field correlator on a disc:
\begin{equation}
  \vev{\psi(x)\sigma^{\ep\ep}(w)}_\zeta
 =2^{-1/2}a\hat{r}(\ep,\zeta)|x-w|^{-1}|w-\bar{w}|^{\frac{3}{8}}.
\end{equation}
 For this correlator to be non-vanishing, we have to identify
 $\sigma^\ep$ with $\bar{\sigma}^{-\ep}$ on the real axis
 up to some constants.

   Another recursion relation can be obtained from the analysis
 of the correlation functions of two spin fields on a disc:
\begin{eqnarray}
 2i\vev{\Theta_{-b/2}^{-\ep,-\ep}(z)\Theta_\alpha^{\ep\ep}(w)}_{u_\zeta}
 &=&
  |z-\bar{w}|^{-4h_{-b/2}-\frac{1}{4}}
  |w-\bar{w}|^{-2h_\alpha+2h_{-b/2}} \times
 \nonumber \\ &&
   \left\{
     -2i\zeta\tilde{C}_+(\alpha) U_+(\alpha-\tfrac{b}{2};u_\zeta)
     \cG_0(p_\alpha,p_{-\frac{b}{2}},-p_\alpha;\eta)
   \right.\nonumber \\ && \left. ~~~
    +\tilde{C}_-(\alpha)U_+(\alpha+\tfrac{b}{2};u_\zeta)
     \cG_3(p_\alpha,p_{-\frac{b}{2}},-p_\alpha;\eta)
   \right\},
\end{eqnarray}
 where $\eta=\left|\frac{z-w}{z-\bar{w}}\right|^2$ and
 the coefficients are determined from the consistency
 with the OPE of two spin fields, as before.
 Note that there seems to be another possibility of writing down the
 solution using $\cG_i(p_\alpha,p_{-b/2},p_\alpha)$
 instead of $\cG_i(p_\alpha,p_{-b/2},-p_\alpha)$.
 However, it will not lead to a recursion relation consistent with
 the analysis of modular property.
 As was discussed in the previous section, it seems that we must
 use appropriate solutions of differential equation according to the total
 chirality.

   Around $\eta\sim1$ the above solution can be expressed as
 a certain linear combination of $\cG_i(p_{-b/2},p_\alpha,-p_\alpha;1-\eta)$.
 Note that all the four of them show up, as opposed to what one
 naively expects as a limit of $\Theta_{-b/2}^{\ep\ep}$
 approaching the boundary.
 According to our observation in the previous section, this is due to
 the fact that we can only fix the total chirality,
 so that the correlator
 $\vev{\Theta_{-b/2}^{\ep\ep}\Theta_\alpha^{\ep\ep}}$
 is actually mixed with
 $\vev{\Theta_{-b/2}^{\ep,-\ep}\Theta_\alpha^{\ep,-\ep}}$
 with equal weights.

   The terms proportional to $\cG_3(p_{-b/2},p_\alpha,-p_\alpha;1-\eta)$
 can be identified with the contribution from the case
 where $\Theta_{-b/2}^{\ep\ep}$ approaches
 the boundary and turns into the boundary identity operator.
 Thus we obtain a recursion relation
\begin{eqnarray}
 \sqrt{2}\lambda^{-2}ir_-(-\ep,\zeta) U_-(\alpha,\ep;u_\zeta)
 &=&
  -2i\zeta\tilde{C}_+(\alpha) U_+(\alpha-\tfrac{b}{2};u_\zeta)
  \frac{\Gamma(b\alpha-\frac{b^2}{2})\Gamma(-b^2)}
  {\sqrt{2}\Gamma(b\alpha-b^2-\frac{1}{2})\Gamma(\frac{1-b^2}{2})}
 \nonumber \\ && \hskip-10mm
  +\tilde{C}_-(\alpha)U_+(\alpha+\tfrac{b}{2};u_\zeta)
  \frac{\Gamma(1+\frac{b^2}{2}-b\alpha)\Gamma(-b^2)}
  {\sqrt{2}\Gamma(\frac{1}{2}-b\alpha)\Gamma(\frac{1-b^2}{2})}.
\label{rcU2}
\end{eqnarray}
 Here a factor $\lambda^{-2}$ was inserted,
 because the solution of the differential equation is actually a mixture
 of correlators as mentioned above and it is not known
 how they are mixed in generic solutions.

   Let us solve the system of two recursion relations.
 We first put the following ansatz:
\begin{eqnarray}
  U_+ &=&
  -2^{-\frac{1}{2}}\pi^{-\frac{1}{2}}
   (\mu\pi\gamma(\tfrac{bQ}{2}))^{\frac{2\alpha-Q}{2b}}
   (\alpha-\tfrac{Q}{2})
  \Gamma(b(\alpha-\tfrac{Q}{2}))
  \Gamma(\tfrac{1}{b}(\alpha-\tfrac{Q}{2}))
  \hat{U}_+, \nonumber \\
  U_- &=&
  \lambda
  2^{-\frac{1}{2}}\pi^{-\frac{1}{2}}
  (\mu\pi\gamma(\tfrac{bQ}{2}))^{\frac{2\alpha-Q}{2b}}
  \Gamma(\tfrac{1}{2}+b(\alpha-\tfrac{Q}{2}))
  \Gamma(\tfrac{1}{2}+\tfrac{1}{b}(\alpha-\tfrac{Q}{2}))
  \hat{U}_-.
\end{eqnarray}
 to simplify the recursion relation into the following form
\begin{eqnarray}
  \hat{U}_-(\alpha-\tfrac{b}{2},-\ep;u_\zeta)
 +\hat{U}_-(\alpha+\tfrac{b}{2},\ep;u_\zeta)
 &=& 2\lambda^{-1}
     \hat{r}(-\ep,\zeta)\mu_B \hat{U}_+(\alpha;u_\zeta)
     \left(\tfrac{1}{2\mu}\cos\tfrac{\pi b^2}{2}\right)^{\frac{1}{2}},
 \nonumber \\
 -\zeta\hat{U}_+(\alpha-\tfrac{b}{2};u_\zeta)
 +\hat{U}_+(\alpha+\tfrac{b}{2};u_\zeta)
 &=& 2\lambda^{-1}
     \hat{r}(-\ep,\zeta)\mu_B
     \hat{U}_-(\alpha,\ep;u_\zeta)
     \left(\tfrac{1}{2\mu}\cos\tfrac{\pi b^2}{2}\right)^{\frac{1}{2}}.
\end{eqnarray}
 The solution for $\zeta=-1$ is given by
\begin{eqnarray}
  \hat{U}_+(\alpha;u_-) &=& \cosh(\pi (2\alpha-Q)u),
 \nonumber \\
  \hat{U}_-(\alpha,\ep;u_-) &=& \cosh(\pi (2\alpha-Q)u),
 \nonumber \\
  \mu_B  &=& \lambda
  \left(\tfrac{2\mu}{\cos(\pi b^2/2)}\right)^{1/2}
   \cosh(\pi u b).
\end{eqnarray}
 together with $\hat{r}(\ep,-)=1$.
 The solution for $\zeta=+1$ becomes
\begin{eqnarray}
  \hat{U}_+(\alpha;u_+) &=& \cosh(\pi (2\alpha-Q)u),
 \nonumber \\
  \hat{U}_-(\alpha,\ep;u_+) &=& \ep \sinh(\pi (2\alpha-Q)u),
 \nonumber \\
  \mu_B  &=& \lambda
  \left(\tfrac{2\mu}{\cos(\pi b^2/2)}\right)^{1/2}
  \sinh(\pi u b).
\end{eqnarray}
 with the condition that $\hat{r}(\ep,+)=-\ep$.
 The above conditions on $\hat{r}$ are met when we assume
 $\psi = -\zeta\bar{\psi}$ on the boundary (real axis).
 The results for $\zeta=-1$ agree with those obtained in \cite{ARS}.

   Our result gives the relation between the label $u$ of
 Cardy states and the boundary cosmological constant $\mu_B$.
 Remarkably, the relation is different according to
 the choice of boundary condition on supercurrent.
 The one-point structure constants for spin fields also differ
 according to $\zeta$.
 Although the corresponding wave function for R Cardy states
 with $\zeta=1$ could not be obtained from the modular property,
 it should be possible to account for this quantity using
 the modular property.

   Due to the subtlety in the correspondence between the solutions
 of differential equation and the correlators, we are left with an
 undetermined constant $\lambda$.
 In later subsection we will see that $\lambda$ should be unity.
 However, we simply set $\lambda=1$ for the time being until we check
 it using the $(3,1)$ degenerate boundary operator.

\paragraph{One-point functions for degenerate boundary states}
   In the same way we can analyze the one-point structure
 constants for boundary states belonging to degenerate representations.
 The main difference as compared to the previous analysis is that
 they have no interpretation in terms of boundary interaction term.
 For bosonic Liouville theory, it was found that the geometry
 of the open worldsheet becomes a pseudosphere\cite{ZZ},
 so that the boundary is infinitely far from generic points in the bulk.
 In this case, disc two-point functions are expected to factorize
 to products of one-point functions in the limit where the two operators
 approach the boundary.

   The recursion relations for degenerate boundary states are
 obtained by a simple modification of (\ref{rcU1}) and (\ref{rcU2}):
\begin{eqnarray} 
 U_-(-\tfrac{b}{2},\ep)U_+(\alpha)
 &=&  C_+(\alpha)U_-(\alpha-\tfrac{b}{2},\ep)
  \frac{\Gamma(b\alpha+\frac{1-b^2}{2})\Gamma(-b^2)}
       {\Gamma(b\alpha-b^2)\Gamma(\frac{1-b^2}{2})}
 ~~~~
 \nonumber \\ &&
  +C_-(\alpha)U_-(\alpha+\tfrac{b}{2},-\ep)
  \frac{\Gamma(\frac{3+b^2}{2}-b\alpha)\Gamma(-b^2)}
       {\Gamma(1-b\alpha)\Gamma(\frac{1-b^2}{2})}
 ,\nonumber \\
 U_-(-\tfrac{b}{2},-\ep)U_-(\alpha,\ep)
 &=&
  -\zeta\tilde{C}_+(\alpha) U_+(\alpha-\tfrac{b}{2})
  \frac{\Gamma(b\alpha-\frac{b^2}{2})\Gamma(-b^2)}
       {\Gamma(b\alpha-b^2-\frac{1}{2})\Gamma(\frac{1-b^2}{2})}
 \nonumber \\ &&
  +\tilde{C}_-(\alpha)U_+(\alpha+\tfrac{b}{2})
  \frac{\Gamma(1+\frac{b^2}{2}-b\alpha)\Gamma(-b^2)}
       {2i\Gamma(\frac{1}{2}-b\alpha)\Gamma(\frac{1-b^2}{2})}.
\label{rcU3}
\end{eqnarray}
 Assuming
\begin{eqnarray}
 U_+(\alpha) &=& \hat{U}_+(\alpha)
  \left\{\mu\pi\gamma(\tfrac{bQ}{2})\right\}^{-\alpha/b}
  \frac{(\alpha-\frac{Q}{2})
        \Gamma(b(\alpha-\frac{Q}{2}))
        \Gamma(\frac{1}{b}(\alpha-\frac{Q}{2}))}
       {(-\frac{Q}{2})\Gamma(-\frac{bQ}{2})\Gamma(-\frac{Q}{2b})},
 \nonumber \\
 U_-(\alpha,+) &=& \hat{U}_-(\alpha)
  \left\{\mu\pi\gamma(\tfrac{bQ}{2})\right\}^{-\alpha/b}
  \frac{\Gamma(\frac{1}{2}+b(\alpha-\frac{Q}{2}))
        \Gamma(\frac{1}{2}+\frac{1}{b}(\alpha-\frac{Q}{2}))}
       {(-\frac{Q}{2})\Gamma(-\frac{bQ}{2})\Gamma(-\frac{Q}{2b})}
\end{eqnarray}
 together with (\ref{SeTf}), they are simplified to the following form:
\begin{eqnarray} 
  \hat{U}_-(-\tfrac{b}{2})\hat{U}_+(\alpha)
 &=& \hat{U}_-(\alpha-\tfrac{b}{2})
     -\zeta \hat{U}_-(\alpha+\tfrac{b}{2}) ,\nonumber \\
  \hat{U}_-(-\tfrac{b}{2})\hat{U}_-(\alpha)
 &=& \hat{U}_+(\alpha-\tfrac{b}{2})
     -\zeta \hat{U}_+(\alpha+\tfrac{b}{2}).
\end{eqnarray}
 Solving them together with those obtained by
 $b\leftrightarrow \tfrac{1}{b}$, we find the following solutions:
\begin{eqnarray}
  \hat{U}_+(\alpha) &=&
  \frac{\sin[\pi br(\alpha-\frac{Q}{2})]
        \sin[\frac{\pi s}{b}(\alpha-\frac{Q}{2})]}
       {\sin\frac{\pi brQ}{2}\sin\frac{\pi sQ}{2b}},
 \nonumber\\
  \hat{U}_-(\alpha) &=& -i^{r+s}
  \frac{\sin\pi r[\frac{1}{2}+b(\alpha-\frac{Q}{2})]
        \sin\pi s[\frac{1}{2}+\frac{1}{b}(\alpha-\frac{Q}{2})]}
       {\sin\frac{\pi brQ}{2}\sin\frac{\pi sQ}{2b}},
\end{eqnarray}
 where $r,s$ are integers whose sum must be even for $\zeta=-1$
 and odd for $\zeta=+1$.
 There exists an ambiguity of $\pm$ sign in front of $\hat{U}_-$
 as is obvious from the structure of the recursion relation.
 The minus sign was chosen from the consistency with the analysis of
 modular property.
 The results for $\zeta=-1$ again agree with \cite{ARS}.
 By making a comparison with the wave functions obtained in the
 previous subsection, we find that the solutions of (\ref{rcU3})
 can all be expressed in terms of them:
\begin{eqnarray}
  U_+(\alpha;(r,s)_\zeta) &=&
  \frac{\Psi_+(-i(\alpha-\tfrac{Q}{2});(r,s)_\zeta)}
       {\Psi_+(\tfrac{iQ}{2};(r,s)_\zeta)},
 \nonumber \\
  U_-(\alpha,+;(r,s)_\zeta) &=&
  \frac{\Psi_-(-i(\alpha-\tfrac{Q}{2});(r,s)_\zeta)}
       {\Psi_+(\tfrac{iQ}{2};(r,s)_\zeta)},
\end{eqnarray}
 if we define the R wave functions for degenerate representations
 as follows:
\begin{eqnarray}
  \Psi_-(p;(r,s)_\zeta) &=& -i^{r+s} 2^{\frac{1}{2}}\pi^{-\frac{1}{2}}
  (\mu\pi\gamma(\tfrac{bQ}{2}))^{-ip/b}
  \Gamma(\tfrac{1}{2}+ipb)\Gamma(\tfrac{1}{2}+\tfrac{ip}{b})
 \nonumber \\ && \hskip10mm \times
  \sin[\pi r(\tfrac{1}{2}+ipb)]
  \sin[\pi s(\tfrac{1}{2}+\tfrac{ip}{b})].
\end{eqnarray}
 The appearance of $\Psi_+(\frac{iQ}{2};(r,s)_\zeta)$ expresses
 that the one-point functions are normalized by the zero-point function
 with the same boundary condition.
 
   Our result for one-point structure constants for degenerate
 representations shows that we should associate the representations
 of NS(R) superalgebras to the boundary states with $\zeta=-1(+1)$.
 This property can also be observed by a detailed analysis of
 the boundary two-point functions.

\subsection{Two-point functions of boundary operators}

   We would like then to find the expression for two-point
 functions of boundary operators on a disc.
 Equivalently, we shall find the boundary reflection coefficients
 $d(\beta|u,u')$ and $\tilde{d}(\beta|u,u')$ defined by
\begin{eqnarray}
  (B_\beta)_{u'_\zeta,u_\zeta}
 &=& d(\beta|u_\zeta,u'_\zeta)(B_{Q-\beta})_{u'_\zeta,u_\zeta},
\nonumber \\
  (\Theta_\beta^\ep)_{u'_{-\zeta},u_\zeta}
 &=& a\tilde{d}(\beta,\ep|u_\zeta,u'_{-\zeta})
  (\Theta_{Q-\beta}^{-\ep})_{u'_{-\zeta},u_\zeta}.
\end{eqnarray}
 Here the boundary states on both sides of the
 boundary operators are expressed by two pairs of $(u,\zeta)$.
 To obtain these coefficients, all we have to do is to study
 the OPE of generic boundary operator with boundary degenerate
 operator $\Theta_{-b/2}^\ep$.
 Note that, as is the case with bosonic Liouville theory with boundary,
 the consistency with fusion algebra requires that the two
 boundary states appearing in the two sides of boundary
 degenerate operators should be related to each other.
 It is expected that the two boundary states $(u,\zeta)$ and
 $(u',\zeta')$ connected by $\Theta_{-b/2}^\ep$ are related via
\begin{equation}
  \pm\, u\pm'u' = \frac{ib}{2},~~~
  \zeta' = -\zeta.
\end{equation}
 The condition is actually more stringent as we will see
 below by a detailed analysis.

   Let us consider the OPE formula
\begin{eqnarray}
\lefteqn{
  \Theta_{-b/2}^\ep(x)_{u''_{-\zeta},u'_\zeta} \times
  B_\beta(y)_{u'_\zeta,u_\zeta}
} \nonumber \\
 &\rightarrow&
  \left\{
  |x-y|^{\frac{b\beta}{2}}
  c_+\cdot\Theta_{\beta-b/2}^\ep(x)
 +|x-y|^{\frac{b(Q-\beta)}{2}}
  ac_-\cdot\Theta_{\beta+b/2}^{-\ep}(x)
  \right\}_{u''_{-\zeta},u_\zeta}
\end{eqnarray}
 and calculate the coefficients $c_\pm(\beta,\ep|u,u',u'';\zeta)$
 using free fields.
 One finds $c_+=1$ as usual, and that $c_-$ is given in terms of
 free field correlators with one boundary screening operator inserted.
 The latter consists of three terms corresponding to three
 boundary segments, and we have to take the sum of these three carefully.
 The result reads
\begin{eqnarray}
\lefteqn{  c_-(\beta,\ep|u,u',u'',\zeta) }
 \nonumber \\
 &=& \frac{-b}{\sqrt{2}r_{\sigma\sigma}(-\ep,\zeta)}
 \left\{
  r_{\sigma\sigma\psi}(-\ep,\zeta)\mu_B
  \frac{\Gamma(1-b\beta)\Gamma(b\beta-\frac{bQ}{2})}
       {\Gamma(1-\frac{bQ}{2})}
 \right. \nonumber \\ && \hskip24mm \left.
 +r_{\sigma\sigma\psi}(-\ep,\zeta)\mu'_B
  \frac{\Gamma(1-b\beta)\Gamma(\frac{bQ}{2})}
       {\Gamma(1+\frac{bQ}{2}-b\beta)}
 +r_{\sigma\psi\sigma}(-\ep,\zeta)\mu''_B
  \frac{\Gamma(b\beta-\frac{bQ}{2})\Gamma(\frac{bQ}{2})}
       {\Gamma(b\beta)}
 \right\} \nonumber \\
 &=& \frac{-b r_{\sigma\psi\sigma}(-\ep,\zeta)}
          {\sqrt{2}\pi r_{\sigma\sigma}(-\ep,\zeta)}
     \Gamma(\tfrac{bQ}{2})\Gamma(1-b\beta)\Gamma(b\beta-\tfrac{bQ}{2})
 \nonumber \\ && \times
 \left\{ -i\ep\mu_B\cos\tfrac{\pi b^2}{2}
         +i\ep\mu'_B\cos\pi(b\beta-\tfrac{b^2}{2})
         +\mu''_B\sin\pi b\beta
 \right\}
\end{eqnarray}
 where the coefficients such as $r_{\sigma\sigma}$ represent
 the prefactors arising in front of free field correlators:
\begin{equation}
\begin{array}{rcl}
  \vev{\sigma^\ep(x)\sigma^\ep(y)}_{\zeta,-\zeta,\zeta}
 &=& r_{\sigma\sigma}(\ep,\zeta)|x-y|^{-1/8}, \\
  \vev{\sigma^\ep(x)\sigma^{-\ep}(y)\psi(z)}_{\zeta,-\zeta,\zeta}
 &=& \tfrac{1}{\sqrt{2}}
     r_{\sigma\sigma\psi}(\ep,\zeta)|x-y|^{3/8}|y-z|^{-1/2}|x-z|^{-1/2},\\
  \vev{\sigma^\ep(x)\psi(y)\sigma^{-\ep}(z)}_{\zeta,-\zeta,\zeta}
 &=& \tfrac{1}{\sqrt{2}}
     r_{\sigma\psi\sigma}(\ep,\zeta)|x-z|^{3/8}|z-y|^{-1/2}|y-z|^{-1/2},\\
  \vev{\psi(x)\sigma^\ep(y)\sigma^{-\ep}(z)}_{\zeta,-\zeta,\zeta}
 &=& \tfrac{1}{\sqrt{2}}
     r_{\psi\sigma\sigma}(\ep,\zeta)|y-z|^{3/8}|x-y|^{-1/2}|x-z|^{-1/2}.
\end{array}
\end{equation}
 There are some relations among them due to the requirement of
 the consistency with cyclic permutations and the analyticity
 in the upper half-plane with respect to the coordinate of $\psi$.
 Fixing them in the following way:
\begin{equation}
  \frac{r_{\sigma\psi\sigma}(\ep,-1)}{r_{\sigma\sigma}(\ep,-1)}  =1,~~~~
  \frac{r_{\sigma\psi\sigma}(\ep,1)}{r_{\sigma\sigma}(\ep,1)}  =-i\ep,
\end{equation}
 and calculating further we obtain
\begin{eqnarray}
 \zeta=-1 \Rightarrow ~~~
  c_-(\beta,\ep|u,u',\zeta)
   &=& \frac{2i\ep b}{\pi}
      (\mu\pi\gamma(\tfrac{bQ}{2}))^{\frac{1}{2}}
      \Gamma(1-b\beta)\Gamma(b\beta-\tfrac{bQ}{2})
 \nonumber \\ && \times
      \sin\tfrac{b\pi}{2}(iu'+iu+\ep\beta)
      \sin\tfrac{b\pi}{2}(iu'-iu+\ep\beta),
 \nonumber \\
 \zeta=+1 \Rightarrow ~~~
  c_-(\beta,\ep|u,u',\zeta)
   &=& \frac{2i b}{\pi}
      (\mu\pi\gamma(\tfrac{bQ}{2}))^{\frac{1}{2}}
      \Gamma(1-b\beta)\Gamma(b\beta-\tfrac{bQ}{2})
 \nonumber \\ && \times
      \cos\tfrac{b\pi}{2}(iu'+iu+\ep\beta)
      \sin\tfrac{b\pi}{2}(iu'-iu+\ep\beta)
\end{eqnarray}
 under the assumption $u''=u'-\frac{i\ep b}{2}$.
 On the other hand, when $u''=u'+\frac{i\ep b}{2}$ the coefficient
 $c_-$ does not reduce to such a simple form.
 This reflects the fact that the boundary states $(u,\zeta)$
 and $(-u,\zeta)$ are not strictly equivalent when $\zeta=1$.
 Hence it is expected that
\begin{equation}
  \left[\Theta_{-b/2}^+\right]_{(u-\frac{ib}{2})_{-\zeta},u_\zeta},~~~~
  \left[\Theta_{-b/2}^-\right]_{(u+\frac{ib}{2})_{-\zeta},u_\zeta}
\label{bSdg}
\end{equation}
 are the only boundary $(2,1)$ degenerate operators
 that are indeed degenerate.
 Combining the reflection equivalence with the OPE formula as in the
 spherical case we obtain the following recursion relations
\begin{eqnarray}
  d(\beta|u_\zeta,u'_\zeta)c_-(Q-\beta,\ep|u,u',\zeta)
 &=& \tilde{d}(\beta-\tfrac{b}{2},\ep|u_\zeta,(u'-\tfrac{ib\ep}{2})_{-\zeta}),
 \nonumber \\
 \tilde{d}(\beta+\tfrac{b}{2},-\ep|u_\zeta, (u'-\tfrac{ib\ep}{2})_{-\zeta})
 c_-(\beta,\ep|u,u',\zeta)
 &=& d(\beta|u_\zeta,u'_\zeta).
\label{rcd1}
\end{eqnarray}

   In the same way, let us consider the OPE
\begin{eqnarray}
\lefteqn{
  \Theta_{-b/2}^\ep(x)_{u''_{-\zeta},u'_\zeta} \times
  \psi B_\beta(y)_{u'_\zeta,u_\zeta}
} \nonumber \\
 &\rightarrow&
  \left\{
  |x-y|^{\frac{b\beta}{2}}
  c'_+\cdot\Theta_{\beta-b/2}^{-\ep}(x)
 +|x-y|^{\frac{b(Q-\beta)}{2}}
  a c'_-\cdot\Theta_{\beta+b/2}^\ep(x)
  \right\}_{u''_{-\zeta},u_\zeta}
\end{eqnarray}
 and calculate the coefficients using free fields.
 One finds
\begin{eqnarray}
  c'_+ &=& \frac{r_{\sigma\sigma\psi}(-\ep,\zeta)}
                {\sqrt{2}r_{\sigma\sigma}(-\ep,\zeta)}
 ~=~ \left\{\frac{-i\ep}{\sqrt{2}}~(\zeta=-1) ~~ ;  ~~
            \frac{1}{\sqrt{2}}~(\zeta=1) \right\}
 \nonumber \\
  c'_- &=&
  \frac{-b}{2\pi r_{\sigma\sigma}(\ep,\zeta)}
  \frac{Q-\beta}{\beta}
  \Gamma(1-b\beta)\Gamma(\tfrac{bQ}{2})\Gamma(b\beta-\tfrac{bQ}{2})
 \nonumber \\ && \times
 \left\{
   r_{\sigma\sigma\psi\psi}(\ep,\zeta)\mu_B
   \cos\tfrac{\pi b^2}{2}
  +r_{\sigma\sigma\psi\psi}(\ep,\zeta)\mu'_B
   \cos\pi(b\beta-\tfrac{b^2}{2})
  -r_{\sigma\psi\sigma\psi}(\ep,\zeta)\mu''_B
   \sin\pi b\beta
 \right\}
 \nonumber \\
 &=&
  \frac{-b(Q-\beta)}{2\pi\beta}
  \Gamma(1-b\beta)\Gamma(\tfrac{bQ}{2})\Gamma(b\beta-\tfrac{bQ}{2})
 \nonumber \\ && \times
 \left\{
   \mu_B \cos\tfrac{\pi b^2}{2}
  +\mu'_B\cos\pi(b\beta-\tfrac{b^2}{2})
  -i\ep \mu''_B\sin\pi b\beta
 \right\}.
\end{eqnarray}
 Here we used the free field correlators
\begin{eqnarray}
\lefteqn{
  \vev{\sigma^\ep(x_1)\sigma^\ep(x_2)\psi(y_1)\psi(y_2)}_{\zeta,-\zeta,\zeta}
}\nonumber \\
 &=& \frac{r_{\sigma\sigma\psi\psi}(\ep,\zeta)}{2}
     \frac{(x_1-y_1)(x_2-y_2)+(x_1-y_2)(x_2-y_1)}
          {|x_1-y_1|^{1/2}|x_1-y_2|^{1/2}|x_2-y_1|^{1/2}|x_2-y_2|^{1/2}
           |x_1-x_2|^{1/8}|y_1-y_2|},
\nonumber \\
\lefteqn{
  \vev{\sigma^\ep(x_1)\psi(y_1)\sigma^\ep(x_2)\psi(y_2)}_{\zeta,-\zeta,\zeta}
}\nonumber \\
 &=& \frac{r_{\sigma\psi\sigma\psi}(\ep,\zeta)}{2}
     \frac{(x_1-y_1)(x_2-y_2)-(x_1-y_2)(y_1-x_2)}
          {|x_1-y_1|^{1/2}|x_1-y_2|^{1/2}|x_2-y_1|^{1/2}|x_2-y_2|^{1/2}
           |x_1-x_2|^{1/8}|y_1-y_2|}.
\end{eqnarray}
 Assuming again $u''=u'-\frac{i\ep b}{2}$, the coefficients
 can be rewritten further:
\begin{eqnarray}
 \zeta=-1 \Rightarrow ~~~
  c'_+(\beta,\ep|u,u',\zeta)
   &=& -i\ep 2^{-\frac{1}{2}},
 \nonumber \\
  c'_-(\beta,\ep|u,u',\zeta)
   &=& -\frac{2^{\frac{1}{2}}b}{\pi}
       \frac{Q-\beta}{\beta}
      (\mu\pi\gamma(\tfrac{bQ}{2}))^{\frac{1}{2}}
      \Gamma(1-b\beta)\Gamma(b\beta-\tfrac{bQ}{2})
 \nonumber \\ && \times
      \cos\tfrac{b\pi}{2}(iu'+iu+\ep\beta)
      \cos\tfrac{b\pi}{2}(iu'-iu+\ep\beta),
 \nonumber \\
 \zeta=+1 \Rightarrow ~~~
  c'_+(\beta,\ep|u,u',\zeta)
   &=& 2^{-\frac{1}{2}},
 \nonumber \\
  c'_-(\beta,\ep|u,u',\zeta)
   &=& \frac{2^{\frac{1}{2}}ib}{\pi}
       \frac{Q-\beta}{\beta}
      (\mu\pi\gamma(\tfrac{bQ}{2}))^{\frac{1}{2}}
      \Gamma(1-b\beta)\Gamma(b\beta-\tfrac{bQ}{2})
 \nonumber \\ && \times
      \sin\tfrac{b\pi}{2}(iu'+iu+\ep\beta)
      \cos\tfrac{b\pi}{2}(iu'-iu+\ep\beta).
\end{eqnarray}
 If we require that the reflection relation holds also for
 descendants $\psi B_\beta$, possibly with the coefficient
 different from that of the primaries
\begin{equation}
  \beta(\psi B_\beta)_{u'_\zeta,u_\zeta}
 = d'(\beta|u_\zeta,u'_\zeta)
  (Q-\beta)(\psi B_{Q-\beta})_{u'_\zeta,u_\zeta},
\end{equation}
 we obtain another set of recursion relations
\begin{eqnarray}
     \tfrac{Q-\beta}{\beta}d'(\beta|u_\zeta,u'_\zeta)
     c'_-(Q-\beta,\ep|u,u',\zeta)
 &=& c'_+(\beta,\ep|u,u',\zeta)
     \tilde{d}(\beta-\tfrac{b}{2},-\ep|u_\zeta,(u'-\tfrac{ib\ep}{2})_{-\zeta}),
 \nonumber \\
      \tfrac{Q-\beta}{\beta}d'(\beta|u_\zeta,u'_\zeta)
      c'_+(Q-\beta,\ep|u,u',\zeta)
 &=&  c'_-(\beta,\ep|u,u',\zeta)
      \tilde{d}(\beta+\tfrac{b}{2},\ep|u_\zeta,(u'-\tfrac{ib\ep}{2})_{-\zeta}).
\label{rcd2}
\end{eqnarray}

   It is straightforward to write down the solutions of the recursion
 relations (\ref{rcd1}) and (\ref{rcd2}) in terms of
 the functions $\bG$ and $\bS$ introduced in \cite{FZZ}.
 They are defined in the following way:
\begin{eqnarray}
  \log \bG(x)&=& \int_0^\infty \frac{dt}{t}\left[
    \frac{e^{-Qt/2}-e^{-xt}}{(1-e^{-bt})(1-e^{-t/b})}
   +\frac{e^{-t}}{2}(\tfrac{Q}{2}-x)^2 +\tfrac{1}{t}(\tfrac{Q}{2}-x) \right] \\
  \log \bS(x) &=&
  \int_0^\infty \frac{dt}{t}\left[\frac{2x-Q}{t}
     -\frac{\sinh[(x-Q/2)t]}{2\sinh[bt/2]\sinh[t/2b]}\right].
\end{eqnarray}
 The function $\bG(x)$ has zeroes at
 $x=-mb-nb^{-1}~(m,n\in {\bf Z})$ and no poles.
 The functions $\bS$ and $\Up$ are expressed in terms of $\bG$:
\begin{equation}
  \bS(x)=\bG(Q-x)/\bG(x),~~~
  \Up(x)=\bG(Q-x)\bG(x).
\end{equation}
 The shift relations for $\bG$ and $\bS$
\begin{equation}
\begin{array}{rcl}
  \bG(x+b)      &=& \bG(x)(2\pi)^{-\frac{1}{2}}b^{\frac{1}{2}-bx}\Gamma(bx),\\
  \bS(x+b)      &=& \bS(x)2\sin(\pi bx),
\end{array}
\begin{array}{rcl}
  \bG(x+\tfrac{1}{b}) &=& \bG(x)
   (2\pi)^{-\frac{1}{2}}b^{\frac{x}{b}-\frac{1}{2}}\Gamma(x/b),\\
  \bS(x+\tfrac{1}{b}) &=& \bS(x)2\sin(\pi x/b)
\end{array}
\end{equation}
 can be used to write down the solutions of the recursion relations.
 If we define the functions $\bG_\NS,\bG_\R$ and $\bS_\NS,\bS_\R$ by
\begin{equation}
\begin{array}{rcl}
  \bG_\NS(x)&=&\bG(\tfrac{x}{2})\bG(\tfrac{x+Q}{2}),\\
  \bS_\NS(x)&=&\bS(\tfrac{x}{2})\bS(\tfrac{x+Q}{2}),
\end{array}
~~
\begin{array}{rcl}
  \bG_\R(x)&=&\bG(\tfrac{x+b}{2})\bG(\tfrac{x+b^{-1}}{2}),\\
  \bS_\R(x)&=&\bS(\tfrac{x+b}{2})\bS(\tfrac{x+b^{-1}}{2}),
\end{array}
\end{equation}
   the solution becomes
\begin{eqnarray}
  d(\beta|u_-,u'_-)
 &=& \frac{(\mu\pi\gamma(\tfrac{bQ}{2})b^{1-b^2})^{\frac{Q-2\beta}{2b}}
           \bG_\NS(Q-2\beta)\bG_\NS(2\beta-Q)^{-1}}
          {\bS_\NS(\beta+iu+iu')
           \bS_\NS(\beta-iu+iu')
           \bS_\NS(\beta+iu-iu')
           \bS_\NS(\beta-iu-iu')},
 \nonumber \\
  d'(\beta|u_-,u'_-)
 &=& \frac{(\mu\pi\gamma(\tfrac{bQ}{2})b^{1-b^2})^{\frac{Q-2\beta}{2b}}
           \bG_\NS(Q-2\beta)\bG_\NS(2\beta-Q)^{-1}}
          {\bS_\R(\beta+iu+iu')
           \bS_\R(\beta-iu+iu')
           \bS_\R(\beta+iu-iu')
           \bS_\R(\beta-iu-iu')},
 \nonumber \\
  \tilde{d}(\beta,\ep|u_-,u'_+)
 &=& \frac{i\ep(\mu\pi\gamma(\tfrac{bQ}{2})b^{1-b^2})^{\frac{Q-2\beta}{2b}}
           \bG_\R(Q-2\beta)\bG_\R(2\beta-Q)^{-1}}
          {\bS_\NS(\beta+iu+i\ep u')
           \bS_\NS(\beta-iu+i\ep u')
           \bS_\R (\beta+iu-i\ep u')
           \bS_\R (\beta-iu-i\ep u')},
 \nonumber \\
  d(\beta|u_+,u'_+)
 &=& \frac{(\mu\pi\gamma(\tfrac{bQ}{2})b^{1-b^2})^{\frac{Q-2\beta}{2b}}
           \bG_\NS(Q-2\beta)\bG_\NS(2\beta-Q)^{-1}}
          {\bS_\R(\beta+iu+iu')
           \bS_\NS(\beta-iu+iu')
           \bS_\NS(\beta+iu-iu')
           \bS_\R(\beta-iu-iu')},
 \nonumber \\
  d'(\beta|u_+,u'_+)
 &=& \frac{-(\mu\pi\gamma(\tfrac{bQ}{2})b^{1-b^2})^{\frac{Q-2\beta}{2b}}
           \bG_\NS(Q-2\beta)\bG_\NS(2\beta-Q)^{-1}}
          {\bS_\NS(\beta+iu+iu')
           \bS_\R(\beta-iu+iu')
           \bS_\R(\beta+iu-iu')
           \bS_\NS(\beta-iu-iu')},
 \nonumber \\
  \tilde{d}(\beta,\ep|u_+,u'_-)
 &=& \frac{i\ep(\mu\pi\gamma(\tfrac{bQ}{2})b^{1-b^2})^{\frac{Q-2\beta}{2b}}
           \bG_\R(Q-2\beta)\bG_\R(2\beta-Q)^{-1}}
          {\bS_\NS(\beta+iu'-i\ep u)
           \bS_\NS(\beta-iu'-i\ep u)
           \bS_\R (\beta+iu'+i\ep u)
           \bS_\R (\beta-iu'+i\ep u)}.
 \nonumber \\
\end{eqnarray}
 They satisfy the unitarity condition and are consistent with
 the equivalence of boundary states $u$ and $-u$ when $\zeta=-1$.
 Note also that the structure of poles of these quantities implies
 that we should identify $\zeta=-1(+1)$ boundary states with
 the representations for NS(R) superalgebras.

   From the above result one can easily read off that $d$ and $d'$
 differ, and $\tilde{d}$ does depend on $\ep$.
 This means that the reflection coefficients of boundary operators
 differ for each operators in a single supermultiplet at least if
 the transformation law is defined in a naive way.
 As a consequence, it follows that the supersymmetry transformation
 and the reflection do not commute.
 However, from the representation theory involving degenerate
 representations it is natural since, if we assume (\ref{bSdg}),
 it follows that different operators in a single (degenerate)
 supermultiplet connect two boundary states in a different way.
 As a simple example, let us consider how the two boundary states
 on the two sides of $B_{-b}$ or $\psi B_{-b}$ are related to
 each other.
 If these operators are regarded as created by multiplying two
 degenerate operators $\Theta_{-b/2}^\ep$, the only possibilities are
\begin{equation}
  \left[B_{-b}\right]_{(u\pm ib)_\zeta,u_\zeta},~~~
  \left[\psi B_{-b}\right]_{u_\zeta,u_\zeta}.
\end{equation}
 This is reasonable if we make comparison with the OPE relations
\begin{equation}
\begin{array}{rcl}
  \left[V_{-b}\right]\times
  \left[V_\alpha\right] &\rightarrow&
  \left[V_{\alpha-b}\right]
 +\left[\psi\bar{\psi}V_{\alpha}\right]
 +\left[V_{\alpha+b}\right],\\
  \left[\psi\bar{\psi}V_{-b}\right]\times
  \left[V_\alpha\right] &\rightarrow&
  \left[\psi\bar{\psi}V_{\alpha-b}\right]
 +\left[V_{\alpha}\right]
 +\left[\psi\bar{\psi}V_{\alpha+b}\right].
\end{array}
\end{equation}
 The generalization of this argument to higher degenerate representation
 is straightforward.
 Let us denote boundary degenerate operators as
 $B_{-kb-hb^{-1}},\psi B_{-kb-hb^{-1}}$ or $\Theta^\ep_{-kb-hb^{-1}}$
 using two non-negative half-integers $k,h$.
 Then the two boundary states with labels $u$ and $u'$
 are related via
\begin{equation}
  u'=u+i(r-k)b+i(s-h)b^{-1},~~
 (0 \le r \le 2k,~~
  0 \le s \le 2h)
\end{equation}
 where $r+s$ must be even for $B_{-kb-hb^{-1}},~\Theta^+_{-kb-hb^{-1}}$
 and odd for $\psi B_{-kb-hb^{-1}},~ \Theta^-_{-kb-hb^{-1}}$.
 Of course this is merely a conjecture, and the consistency
 should be proven by a more detailed analysis of this model.

\paragraph{Density of open string states}

   The reflection coefficients can be regarded as phase shifts,
 so are related to the density of certain open string states through the
 formulae of the following form:
\begin{eqnarray}
  \frac{1}{2\pi i}\frac{d}{ds}\log d(\tfrac{Q}{2}+is|u_\zeta,u'_\zeta)
  &=& \rho(s|u_\zeta,u'_\zeta),
 \nonumber \\
  \frac{1}{2\pi i}\frac{d}{ds}\log d'(\tfrac{Q}{2}+is|u_\zeta,u'_\zeta)
  &=& \rho'(s|u_\zeta,u'_\zeta),
 \nonumber \\
  \frac{1}{2\pi i}\frac{d}{ds}\log
  \tilde{d}(\tfrac{Q}{2}+is,\ep|u_\zeta,u'_\zeta)
  &=& \tilde{\rho}(s,\ep|u_\zeta,u'_\zeta).
\end{eqnarray}
 Using the integral expression for $\bS$ and
 discarding terms independent of the labels $u$ and $u'$,
 some of them are given by
\begin{eqnarray}
  \rho(s|u_-,u'_-)
 &\sim&
  2\int_{-\infty}^\infty dp
   \frac{e^{2\pi ips}\cosh(\pi Qp)\cos(2\pi up)\cos(2\pi u'p)}
        {\sinh(2\pi bp)\sinh(2\pi p/b)}
 \nonumber \\
 &=&
  \int_{-\infty}^\infty\frac{dp}{\pi}e^{2\pi ips}
  \left[\Psi_+(p;u_-)\Psi^\dag_+(p;u'_-)
       +\Psi_-(p;u_-)\Psi_-^\dag(p;u_-)\right],
 \nonumber\\
  \rho'(s|u_-,u'_-)
 &\sim&
  2\int_{-\infty}^\infty dp
   \frac{e^{2\pi ips}\cosh\pi(b-b^{-1})p\cos(2\pi up)\cos(2\pi u'p)}
        {\sinh(2\pi bp)\sinh(2\pi p/b)}
 \nonumber\\
 &=&
  \int_{-\infty}^\infty\frac{dp}{\pi}e^{2\pi ips}
  \left[\Psi_+(p;u_-)\Psi_+^\dag(p;u'_-)
       -\Psi_-(p;u_-)\Psi_-^\dag(p;u_-)\right].
\end{eqnarray}
 This agrees with the analysis of modular property since the annulus
 partition function bounded by two Cardy states $u_-$ and $u'_-$
 is given by
\begin{eqnarray}
  Z_{u_-,u'_-} &=&
   \int_{-\infty}^\infty \frac{dp}{2\pi}  \left[
    2\Psi_+(p;u_-)\Psi_+^\dag(p;u'_-)\chi_{p(\NS)}^+(\tau_c)
   +\sqrt{2}\Psi_-(p;u_-)\Psi_-^\dag(p;u'_-)\chi_{p(\R) }^+(\tau_c)
  \right]
  \nonumber \\ &=&\frac{1}{2}
  \int_{-\infty}^\infty ds\left[
    \rho (s|u_-,u'_-)\left\{\chi_{s(\NS)}^++\chi_{s(\NS)}^-\right\}(\tau_o)
  \right. \nonumber \\ && \left. ~~~~~~~~~~~
   +\rho'(s|u_-,u'_-)\left\{\chi_{s(\NS)}^+-\chi_{s(\NS)}^-\right\}(\tau_o)
  \right].
\end{eqnarray}
 The other quantities can be written as
\begin{eqnarray}
  \rho(s|u_+,u'_+)
 &=&
  \int_{-\infty}^\infty\frac{dp}{\pi}e^{2\pi ips}
  \left[\Psi_+(p;u_+)\Psi^\dag_+(p;u'_+)
       +\Psi_-(p;u_+)\Psi_-^\dag(p;u'_+)\right],
 \nonumber\\
  \rho'(s|u_+,u'_+)
 &=&
  \int_{-\infty}^\infty\frac{dp}{\pi}e^{2\pi ips}
  \left[\Psi_+(p;u_+)\Psi_+^\dag(p;u'_+)
       -\Psi_-(p;u_+)\Psi_-^\dag(p;u'_+)\right],
 \nonumber \\
  \tilde{\rho}(s,\ep|u_-,u'_+)
 &=&
  \int_{-\infty}^\infty\frac{dp}{\pi}e^{2\pi ips}
  \left[\Psi_+(p;u_-)\Psi^\dag_+(p;u'_+)
    +\ep\Psi_-(p;u_-)\Psi_-^\dag(p;u'_+)\right],
 \nonumber\\
  \tilde{\rho}(s,\ep|u_+,u'_-)
 &=&
  \int_{-\infty}^\infty\frac{dp}{\pi}e^{2\pi ips}
  \left[\Psi_+(p;u_+)\Psi_+^\dag(p;u'_-)
    +\ep\Psi_-(p;u_+)\Psi_-^\dag(p;u'_-)\right].
\end{eqnarray}
 if we assume that the R wave function for $\zeta=1$ is given by
\begin{equation}
  \Psi_-(p;u_+) = i2^{-\frac{1}{2}}\pi^{-\frac{1}{2}}
  (\mu\pi\gamma(\tfrac{Q}{2}))^{-ip/b}
  \Gamma(\tfrac{1}{2}+ipb)\Gamma(\tfrac{1}{2}+\tfrac{ip}{b})
  \sin(2\pi pu).
\label{RPs+}
\end{equation}
 Hence, in all the cases the phase shifts give the density of open
 string states with definite worldsheet fermion number.
 This also suggests that the wave function for R Cardy states with
 $u_+$ is given by (\ref{RPs+}), in consistency with the analysis
 of one-point structure constants.

\paragraph{Bulk and boundary cosmological constants}

   As a further check, let us consider the OPE of boundary
 operators involving $(3,1)$ degenerate representation.
 Above all, the OPE coefficients involving them depend on
 both of $\mu$ and $\mu_B$.
 The consistency with the previous arguments fixes the constant
 $\lambda$ which was left undetermined.

   Let us consider the OPE relation
\begin{eqnarray}
\lefteqn{
  \left[B_{-b}\right]_{u''_\zeta,u'_\zeta}
  \left[B_\beta\right]_{u'_\zeta,u_\zeta}
}\nonumber \\ & \rightarrow &
  \hat{c}_+\left[B_{\beta-b} \right]_{u''_\zeta,u_\zeta}
 +\hat{c}_0\left[\psi B_\beta\right]_{u''_\zeta,u_\zeta}
 +\hat{c}_-\left[B_{\beta+b} \right]_{u''_\zeta,u_\zeta}
\end{eqnarray}
 with $u''=u'+ ib$, and use $\hat{c}_\pm$ to derive another
 recursion relation for boundary reflection coefficient.
 Here again $\hat{c}_+=1$, so we concentrate on the calculation
 of $\hat{c}_-$.
 There are largely two contributions to $\hat{c}_-$,
 which are proportional to $\mu$ and $\mu_B^2$, respectively.
 Calculating first the contribution proportional $\mu$
 using free fields, one finds
\begin{equation}
   2\mu b^2\zeta\int d^2z|z|^{-2b\beta}|1-z|^{2b^2}|z-\bar{z}|^{-b^2-1}
  = -2\mu b^2\zeta I_0 \sin(\pi b^2)\sin^2(\pi b\beta)
\end{equation}
 where $I_0$ is given by
\begin{equation}
  I_0 = -\frac{\gamma(\tfrac{bQ}{2})}
             {2\pi\sin\pi b^2}
  \Gamma(1-b\beta)\Gamma(1-\tfrac{bQ}{2}-b\beta)
  \Gamma(b\beta)\Gamma(b\beta-\tfrac{bQ}{2}).
\end{equation}
 There are six contributions proportional to $\mu_B^2$, since
 there are six ways of inserting two boundary screening operators
 onto the boundary divided into three segments.
 Restoring $\lambda$, they are summarized into the following form
\begin{eqnarray}
&&b^2 I_0\left\{
   -\mu_B^2 \sin(\pi b^2)\cos\tfrac{\pi b^2}{2}
   -{\mu'_B}^2\sin(\pi b\beta)\cos\pi(b\beta-\tfrac{b^2}{2})
   +{\mu''_B}^2\sin(\pi b\beta)\cos\pi(b\beta+\tfrac{b^2}{2})  
 \right\}
 \nonumber \\
&&+2b^2 I_0\sin\tfrac{\pi b^2}{2}\left\{
    \mu_B\mu'_B\cos\tfrac{\pi b^2}{2}\cos\pi(b\beta-\tfrac{b^2}{2})
   -\mu_B\mu''_B\cos\tfrac{\pi b^2}{2}\cos\pi(b\beta+\tfrac{b^2}{2})
 \right. \nonumber \\ && \left. \hskip30mm
   +\mu'_B\mu''_B\cos\pi(b\beta-\tfrac{b^2}{2})\cos\pi(b\beta+\tfrac{b^2}{2})
 \right\}
 \nonumber \\
 & & ~~(\zeta=-1) \nonumber \\
 &=& -2\mu\lambda^2b^2I_0 \sin(\pi b^2)
    \left\{
    \sin^2(\pi b\beta)
   +4\sin\tfrac{b\pi}{2}(iu'+iu+\beta) 
     \sin\tfrac{b\pi}{2}(iu'-iu+\beta) 
 \right. \nonumber \\ && \left.
  \hskip50mm \times
     \cos\tfrac{b\pi}{2}(iu'+iu+\beta+b) 
     \cos\tfrac{b\pi}{2}(iu'-iu+\beta+b) 
    \right\}
 \nonumber \\
 & & ~~(\zeta=1) \nonumber \\
 &=& +2\mu\lambda^2b^2I_0 \sin(\pi b^2)
    \left\{
    \sin^2(\pi b\beta)
   -4\cos\tfrac{b\pi}{2}(iu'+iu+\beta) 
     \sin\tfrac{b\pi}{2}(iu'-iu+\beta) 
 \right. \nonumber \\ && \left.
  \hskip50mm \times
     \sin\tfrac{b\pi}{2}(iu'+iu+\beta+b) 
     \cos\tfrac{b\pi}{2}(iu'-iu+\beta+b) 
    \right\}.
\end{eqnarray}
 One can see that when $\lambda=1$ there are some cancellation
 between the contributions from the bulk and the boundary,
 and the result for $\hat{c}_-$ yields a recursion relation
 for $d(\beta|u_\zeta,u'_\zeta)$ consistent with the analysis
 using $\Theta_{-b/2}^\ep$.
 
\section{Concluding remarks}

   In this paper the $N=1$ super Liouville theory was analyzed
 on a sphere and on a disc.
 The analysis was based on the approach developed in
 \cite{T,ZZ2,FZZ,ZZ} for bosonic Liouville theory.
 Various quantities were obtained in a form very similar to the
 bosonic case.
 However, there are some new features in $N=1$ theory largely
 due to the fermionic nature of screening operators.
 As one of the consequences, the reflection of spin fields
 are always accompanied by the flip of the chirality.
 We also presented all the solutions of differential equations
 for four-point functions containing one degenerate spin operator
 $\Theta_{-b/2}^{\ep}$.
 For four-point functions of four spin fields, the differential
 equation become of the fourth order and there are therefore four
 independent solutions in apparent contradiction with the assumption
 that the product of $\Theta_{-b/2}^{\ep}$ with any operators
 is decomposed into two discrete terms.
 However, our solutions obey a special transformation property
 under the change of basis so that the crossing symmetric combination
 of the left and the right sectors can be identified with
 a particular sum of four-point functions of spin fields.

   Contrarily to the bosonic case, the analysis of
 modular property of annulus partition functions was not
 sufficient to obtain all the wave functions that define Cardy
 states.
 It was also found that the two ways of putting boundary condition
 on supercurrent leads to two boundary states which differ in
 quite a non-trivial way.
 Indeed, we were unable to find the R Cardy states with $\zeta=1$
 from the modular analysis.
 We found the wave functions for them through the analysis of
 disc one-point functions and found the correspondence between
 $\zeta=-1(+1)$ boundary states and the representations
 of NS(R) superalgebras.

   The two-point functions of boundary operators were also obtained
 and the density of open string states which can be read off from
 them were shown to be consistent with the analysis of modular property.
 Remarkably, the reflection coefficients for boundary operators
 depend on the label of boundary states in such a way that they
 are different for different components in the same supermultiplet.
 To understand this we need a more detailed analysis of the property
 of boundary operators.

   Our analysis have shown that the non-compact superconformal theory
 with boundary can be analyzed using the techniques developed in the
 analysis of bosonic theory, if an appropriate care is taken.
 It would then be interesting to analyze similar superconformal
 theories or those with higher worldsheet supersymmetry in the same way.

\paragraph{Note added}
   For disc one-point functions, after the submission of this
 paper we were informed of the preceding analysis \cite{ARS}
 which covers the $\zeta=-1$ case of our result.

\paragraph{Acknowledgments}
   The authors thank C. Ahn, T. Eguchi, K. Ito, I. Kishimoto,
 C. Rim, M. Stanishkov and Al.B. Zamolodchikov
 for discussions and comments.
 The works of the authors were supported in part by JSPS research
 fellowships for young scientists.

\newpage

\begin{center}{\sc References}\end{center}\par

\list{[\arabic{enumi}]}
     {\settowidth\labelwidth{[99]}\leftmargin\labelwidth
      \advance\leftmargin\labelsep\usecounter{enumi}}
\def\newblock{\hskip .11em plus .33em minus .07em}
\sloppy\clubpenalty4000\widowpenalty4000
\sfcode`\.=1000\relax
\let\endthebibliography=\endlist


\bibitem{FZZ} V. Fateev, A. Zamolodchikov and Al. Zamolodchikov,
    {\sl ``Boundary Liouville Field Theory I.
           Boundary State and Boundary Two-point Function''},
    {\tt hep-th/0001012}.

\bibitem{ZZ} A. Zamolodchikov and Al. Zamolodchikov,
    {\sl ``Liouville field theory on a pseudosphere''},
    {\tt hep-th/0101152}.

\bibitem{H} K. Hosomichi,
    {\sl ``Bulk-Boundary Propagator in Liouville Theory on a Disc''},
    JHEP {\bf 0111}, 044 (2001),
    {\tt hep-th/0108093}.

\bibitem{PT} B. Ponsot and J. Teschner,
    {\sl ``Boundary Liouville field theory: Boundary three point function''},
    Nucl. Phys. B {\bf 622}, 309 (2002),
    {\tt hep-th/0110244}.

\bibitem{GKS} A. Giveon, D. Kutasov and A. Schwimmer,
    {\sl ``Comments on D-branes in $AdS_3$''},
    Nucl. Phys. B {\bf 615}, 133 (2001),
    {\tt hep-th/0106005}.

\bibitem{PS} A. Parnachev and D.A. Sahakyan,
    {\sl ``Some Remarks on D-branes in $AdS_3$''},
    JHEP {\bf 0110}, 022 (2001),
    {\tt hep-th/0109150}.

\bibitem{LOP} P. Lee, H. Ooguri and J.w. Park,
    {\sl ``Boundary States for $AdS_2$ Branes in $AdS_3$''},\\
    {\tt hep-th/0112188}.

\bibitem{PST} B. Ponsot, V. Schomerus and J. Teschner,
    {\sl ``Branes in the Euclidean $AdS_3$''},\\
    {\tt hep-th/0112198}.

\bibitem{Arvis} J.F. Arvis,
    {\sl ``Classical Dynamics of The Supersymmetric Liouville Theory''},
    Nucl. Phys. B {\bf 212}, 151 (1983);
    {\sl ``Spectrum of The Supersymmetric Liouville Theory''},
    Nucl. Phys. B {\bf 218}, 309 (1983).

\bibitem{DHoker} E. D'Hoker,
    {\sl ``Classical And Quantal Supersymmetric Liouville Theory''},
    Phys. Rev. D {\bf 28}, 1346 (1983).

\bibitem{Babelon} O. Babelon,
    {\sl ``Monodromy Matrix and its Poisson Brackets
           in Supersymmetric Liouville String Theory''},
    Phys. Lett. B {\bf 141}, 353 (1984);
    {\sl ``Construction of the Quantum Supersymmetric Liouville Theory
           for String Models''},
    Nucl. Phys. B {\bf 258}, 680 (1985).

\bibitem{ZP} A.B. Zamolodchikov and R.G. Pogosian,
    {\sl ``Operator Algebra in Two-Dimensional Superconformal Field Theory''},
    Sov. J. Nucl. Phys. {\bf 47}, 929 (1988).

\bibitem{ACDH} E. Abdalla, M.C. Abdalla, D. Dalmazi and K. Harada,
    {\sl ``Correlation functions in super Liouville theory,''}
    Phys. Rev. Lett. {\bf 68}, 1641 (1992),
    {\tt hep-th/9108025}.

\bibitem{dFK} P. Di Francesco and D. Kutasov,
    {\sl ``World Sheet and Space-Time Physics
           in Two Dimensional (Super)String Theory,''}
    Nucl. Phys. B {\bf 375}, 119 (1992),
    {\tt hep-th/9109005}.

\bibitem{AD} K.i. Aoki and E. D'Hoker,
    {\sl ``On the Liouville approach to correlation functions
           for 2-D quantum gravity''},
    Mod. Phys. Lett. A {\bf 7}, 235 (1992),
    {\tt hep-th/9109024};
    {\sl ``Correlation functions of minimal models coupled to
           two-dimensional quantum supergravity''},
    Mod. Phys. Lett. A {\bf 7}, 333 (1992),
    {\tt hep-th/9109025}.

\bibitem{DA} D. Dalmazi and E. Abdalla,
    {\sl ``Correlators in noncritical superstrings including
           the spinor emission vertex,''}
    Phys. Lett. B {\bf 312}, 398 (1993),
    {\tt hep-th/9302032}.

\bibitem{Poghosian} R.H. Poghosian,
    {\sl ``Structure constants in the $N=1$ super-Liouville field theory''}
    Nucl. Phys. B {\bf 496}, 451 (1997),
    {\tt hep-th/9607120}.

\bibitem{RS} R.C. Rashkov and M. Stanishkov,
    {\sl ``Three-point correlation functions
           in $N=1$ Super Liouville Theory''},
    Phys. Lett. B {\bf 380}, 49 (1996),
    {\tt hep-th/9602148}.

\bibitem{ARS} C. Ahn, C. Rim and M. Stanishkov,
    {\sl ``Exact One-Point Function of $N=1$ super-Liouville
           Theory with Boundary''},
    {\tt hep-th/0202043}.

\bibitem{Polyakov2} A.M. Polyakov,
    {\sl ``Quantum Geometry of Fermionic Strings,''}
    Phys. Lett. B {\bf 103}, 211 (1981).

\bibitem{DHK} J. Distler, Z. Hlousek and H. Kawai,
    {\sl ``SuperLiouville Theory as a Two-Dimensional, Superconformal
           Supergravity Theory,''}
    Int. J. Mod. Phys. A {\bf 5}, 391 (1990).

\bibitem{CKT} S. Chaudhuri, H. Kawai and S.H. Tye,
    {\sl ``Path Integral Formulation Of Closed Strings,''}
    Phys. Rev. D {\bf 36}, 1148 (1987).

\bibitem{GS} M.B. Green and N. Seiberg,
    {\sl ``Contact Interactions in Superstring Theory,''}
    Nucl. Phys. B {\bf 299}, 559 (1988).

\bibitem{DS} M. Dine and N. Seiberg,
    {\sl ``Microscopic Knowledge from Macroscopic Physics in String Theory,''}
    Nucl. Phys. B {\bf 301}, 357 (1988).

\bibitem{GL} M. Goulian and M. Li,
    {\sl ``Correlation functions in Liouville theory,''}
    Phys. Rev. Lett. {\bf 66}, 2051 (1991).

\bibitem{BKT} M.A. Bershadsky, V.G. Knizhnik and M.G. Teitelman,
    {\sl ``Superconformal Symmetry In Two Dimensions,''}
    Phys. Lett. B {\bf 151}, 31 (1985).

\bibitem{FQS} D. Friedan, Z.a. Qiu and S.H. Shenker,
    {\sl ``Superconformal Invariance in Two-Dimensions
           and the Tricritical Ising Model,''}
    Phys. Lett. B {\bf 151}, 37 (1985).

\bibitem{Nam} S.k. Nam,
    {\sl ``The Kac Formula For the $N=1$ and the $N=2$
           Superconformal Algebras,''}
    Phys. Lett. B {\bf 172}, 323 (1986).

\bibitem{T} J. Teschner,
    {\sl ``On the Liouville three point function''},
    Phys. Lett. B {\bf 363}, 65 (1995), {\tt hep-th/9507109}.

\bibitem{DO} H. Dorn and H. J. Otto,
    {\sl ``Two and three point functions in Liouville theory''},
    Nucl. Phys. B {\bf 429}, 375 (1994), {\tt hep-th/9403141}.

\bibitem{ZZ2} A. Zamolodchikov and Al. Zamolodchikov,
    {\sl ``Structure constants and conformal bootstrap
           in Liouville field theory''},
    Nucl. Phys. B {\bf 477}, 577 (1996), {\tt hep-th/9506136}.

\bibitem{BPZ} A.A. Belavin, A.M. Polyakov and A.B. Zamolodchikov,
    {\sl ``Infinite Conformal Symmetry In Two-Dimensional
           Quantum Field Theory''},
    Nucl. Phys. B {\bf 241}, 333 (1984).

\bibitem{GZ} S. Ghoshal and A.B. Zamolodchikov,
    {\sl ``Boundary S matrix and boundary state
           in two-dimensional integrable quantum field theory''},
    Int. J. Mod. Phys. A {\bf 9}, 3841 (1994),
    Erratum-ibid. A {\bf 9}, 4353 (1994),
    {\tt hep-th/9306002}.

\bibitem{N} R.I. Nepomechie,
    {\sl ``The boundary supersymmetric sine-Gordon model revisited''},
    Phys. Lett. B {\bf 509}, 183 (2001),
    {\tt hep-th/0103029}.

\end{document}